\documentclass[preprint,3p]{elsarticle}

\usepackage{amssymb}

\usepackage{graphicx}
\usepackage{color}

\begin{document}

\begin{frontmatter}

\title{WIMP Dark Matter Direct-Detection Searches in Noble Gases}
 \author{Laura Baudis}
 \address{Physik Institut, University of Zurich, Winterthurerstrasse 190, 8057 Zurich}
 
\author{}
\address{}

\date{\today}

\begin{abstract} 

Cosmological observations and the dynamics of the Milky Way provide ample evidence for an invisible and dominant mass component. 
This so-called dark matter could be made of new, colour and charge neutral particles, which were  non-relativistic when they decoupled from ordinary  matter 
in the early universe.  Such weakly interacting massive particles (WIMPs) are predicted to have a non-zero coupling to baryons and could 
be detected via their collisions with atomic nuclei in ultra-low background, deep underground detectors. Among these, detectors based on liquefied noble gases have demonstrated tremendous discovery potential over the last decade. After briefly introducing the phenomenology of direct dark matter detection, I will review the main properties of liquefied argon and xenon as WIMP targets and discuss sources of background. I will then describe existing and planned argon and xenon detectors that employ the so-called single- and dual-phase detection techniques, addressing their complementarity and science reach.

\end{abstract}

\end{frontmatter}

\section{Introduction}

The motivation to build detectors that can directly detect particle dark matter in the laboratory comes from our current understanding of the observable universe.
Cosmological observations ranging from the measured abundance of primordial elements to the precise  
mapping of anisotropies in the cosmic microwave background, to the study of the distribution of matter on galactic, extragalactic and the largest observed scales, to observations of high-redshift supernovae, have led to a so-called standard model of cosmology. 
In this  model, our universe is spatially flat and composed of $\sim$5\% atoms, $\sim$27\% dark matter and $\sim$68\% dark energy \cite{Ade:2013zuv}.  
Understanding the nature of dark matter  poses a significant challenge to astroparticle and particle physics, for its solution 
may involve new particles with masses and cross sections characteristic of the electroweak scale.
Such weakly interactive massive particles {\small (WIMPs)}, which would have been in thermal equilibrium with quarks and leptons in the hot 
early universe, and decoupled when they were non-relativistic, represent a generic class of dark matter candidates~\cite{Lee:1977ua}. 
Concrete examples are the lightest superpartner in supersymmetry with R-parity conservation \cite{Jungman:1995df}, and the lightest Kaluza-Klein particle, for instance  the first excitation of  the hypercharge gauge boson, in theories with universal extra dimensions  \cite{Hooper:2007qk}. Perhaps the most intriguing aspect of the {\small WIMP} hypothesis is the fact that it is testable by experiment. {\small WIMPs} with masses around the TeV scale are within reach of high-energy colliders and of direct and indirect dark matter searches \cite{Bertone:2004pz}.  Axions and axion-like particles (ALPs), produced non-thermally in the early universe,  represent another class of well-motivated dark matter candidates \cite{Raffelt:2002zz}. They can be searched for by exploiting their predicted couplings to photons and electrons, where the most successful technique to search for dark matter axions is the so-called axion haloscope, that relies on the axion to two-photon coupling \cite{Baker:2013zta,Rybka:2014xca}.  In this review I will focus on the search for WIMPs with low-background detectors operated underground.

\section{Direct detection of WIMPs}

Dark matter particles might be detected via their elastic collisions with atomic nuclei in earthbound, low-background detectors \cite{Goodman:1984dc}. 
The differential rate for  elastic scattering can be expressed as  \cite{Lewin:1995rx}:

\begin{equation}
\frac {dR}{dE_R}=N_{T}
\frac{\rho_{dm}}{m_{W}}
                    \int_{v_{\rm min}}^{v_{\rm max}} \,d \vec{v}\,f(\vec v)\,v
                     \,\frac{d\sigma}{d E_R}\,  
\label{eq1}
\end{equation}

\noindent
where $N_T$ is the number of target nuclei, $\rho_{dm}$ is the local dark matter density in the galactic halo, $m_W$ is the {\small WIMP} mass,
$\vec v$ and $f(\vec v)$ are the {\small WIMP} velocity and velocity distribution function  in the Earth frame and ${d\sigma}/{d E_R}$ is the {\small WIMP}-nucleus differential cross section. The nuclear recoil energy is $E_R={{m_{\rm r}^2}}v^2(1-\cos \theta)/{m_N}$, where $\theta$ is the  scattering angle in the  center-of-mass frame, $m_N$ is the nuclear mass and $m_{\rm r}$ is the reduced mass. The minimum velocity is defined as  $v_{\rm min} = (m_N E_{th}/2m_{\rm r}^2)^{\frac{1}{2}}$, where $E_{th}$ is the energy threshold of the detector, and $v_{\rm max}$ is the escape velocity in the Earth frame.  The simplest galactic model assumes a Maxwell-Boltzmann distribution for the WIMP velocity in the galactic rest frame, with a velocity dispersion of $\sigma \approx$ 270\,km\,s$^{-1}$ and an escape velocity of v$_{esc} \approx$ 544\,km\,s$^{-1}$ \cite{Green:2011bv,Smith:2006ym}.  For  direct detection experiments, the mass density  and velocity distribution at a radius around 8\,kpc are most relevant. State of the art, dark-matter-only cosmological simulations of Milky Way-like halos find that the dark matter mass distribution at the solar position is smooth,  with substructures   being far away from the Sun. The local velocity distribution of dark matter particles is likewise found to be smooth, and  close to Maxwellian \cite{Vogelsberger:2008qb}.  The efforts to measure the mean density of dark matter near our Sun are extensively reviewed in \cite{Read:2014qva}: the Milky Way seems to be consistent with featuring a spherical dark matter halo with little flattening in the disc plane, and $\rho_{dm}$=$0.2-0.56$\,GeV\,cm$^{-3}$ \cite{Read:2014qva}. The largest source of uncertainty in $\rho_{dm}$ comes from the baryonic contribution to the local dynamical mass.

The differential cross section for elastic scattering is  traditionally split in two leading components: an effective scalar coupling between the {\small WIMP} and the mass of the nucleus  and  an effective coupling between the spin of the {\small WIMP} and the total angular momentum of the nucleus. In general, the coherent part dominates the interaction (depending however on the characteristics and composition of the dark matter particle) for target masses with A$\geq$30 \cite{Jungman:1995df}. The total cross section  is  the sum of both contributions    $\frac{d\sigma}{d E_R} \propto {\sigma_{SI}^0} F_{SI}^2(E_R) + {\sigma_{SD}^0} F_{SD}^2(E_R)$, where  $\sigma_{SI,SD}^0$ are the spin-independent (SI) and  spin-dependent (SD)  cross sections in the limit of zero momentum transfer and $F_{SI,SD}^2(E_R)$ denote the nuclear form factors, expressed as a function of the recoil energy. These become significant at large WIMP and nucleus masses, leading to a suppression of the differential scattering rate at higher recoil energies. State-of-the art shell-model calculations for spin-dependent scattering were recently performed for most non-zero spin nuclei of interest in direct searches \cite{Menendez:2012tm,Klos:2013rwa}. In addition, in these new calculations the WIMP-nucleus currents are based on chiral effective field theory at the one-body level, and also include the leading long-range two-body currents.  

Recent, more general treatments of the WIMP-nucleon cross section in non-relativistic, effective field theory approaches \cite{Fan:2010gt, Fitzpatrick:2012ix, Hill:2013hoa, Anand:2013yka}, identify a full set of operators that describe the potential interactions of dark matter particles, and relate these to nuclear response functions. These are of great interest because, in principle, more information can be extracted from direct detection experiments in case of a signal detection, assuming a certain variety of targets is employed in the search. 

\begin{figure}[h!]
\begin{minipage}[]{0.5\textwidth}
\begin{center}
\includegraphics[width=0.975\columnwidth,clip=]{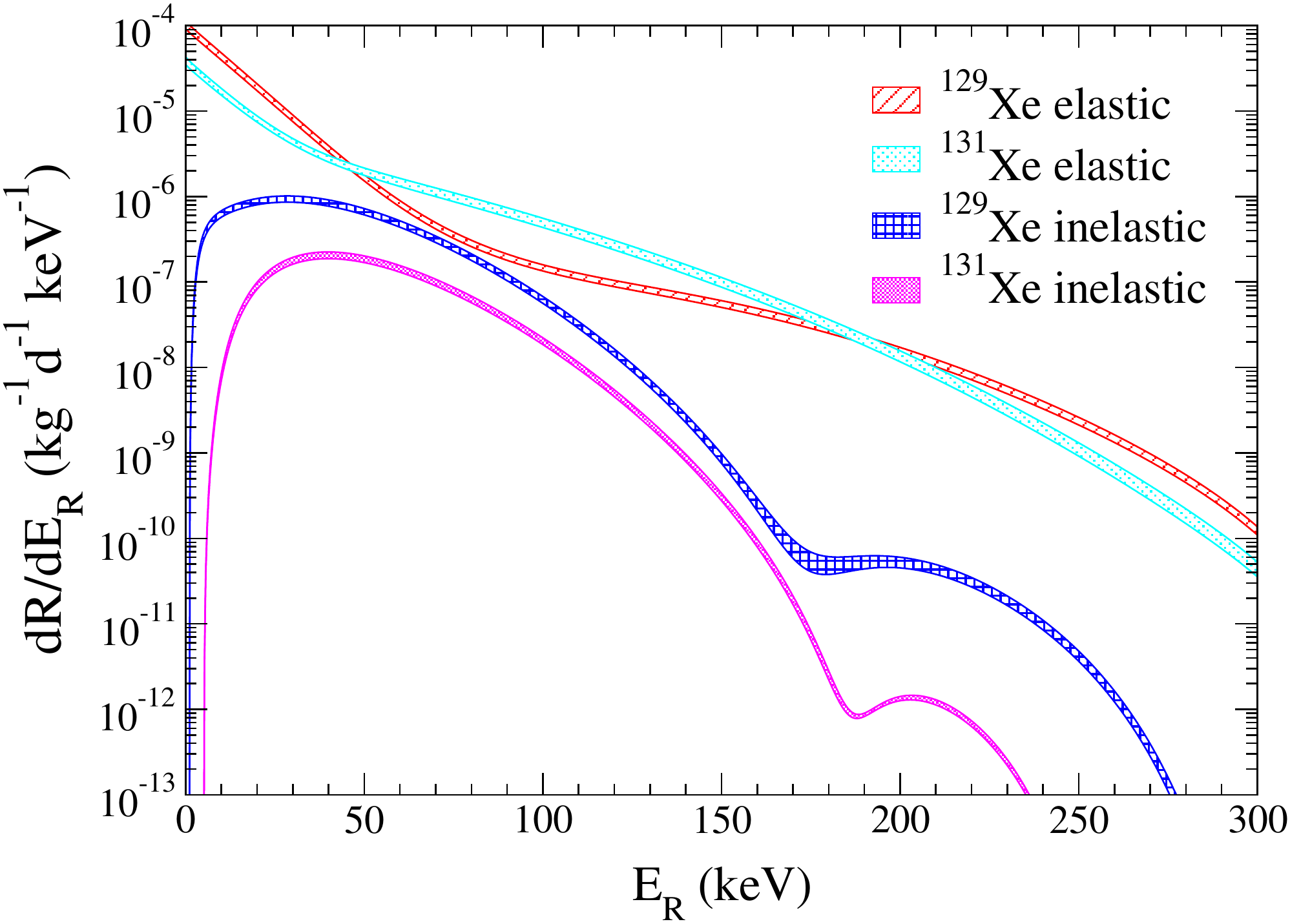}
\end{center}
\end{minipage}
\hfill
\begin{minipage}[]{0.5\textwidth}
\begin{center}
\includegraphics[width=0.975\columnwidth,clip=]{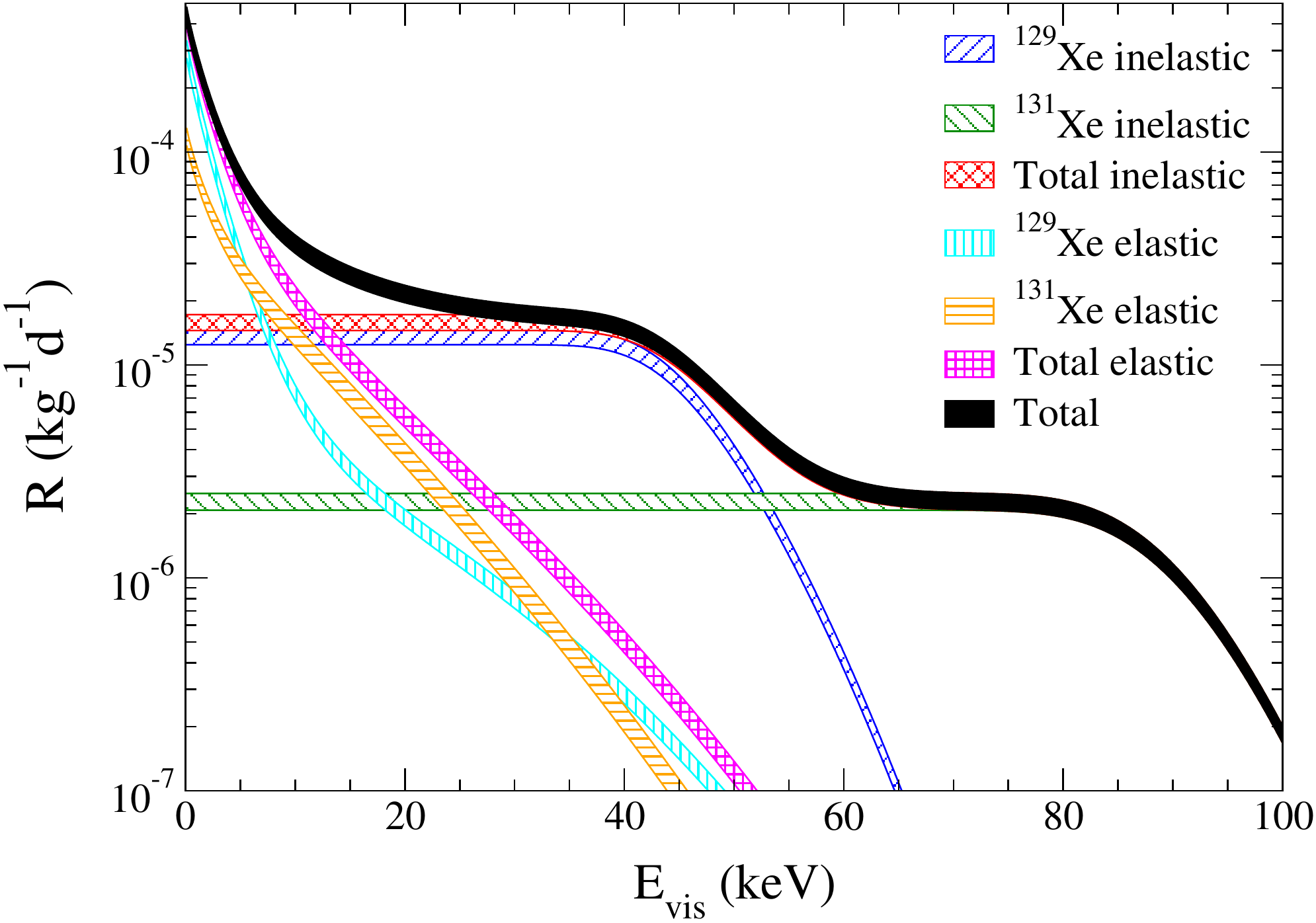}
\end{center}
\end{minipage}
\caption{(Left) Differential nuclear recoil spectra as a function of recoil energy for scattering off $^{129}$Xe and $^{131}$Xe.  Both elastic and inelastic
recoil spectra are shown for comparison,  the bands include the uncertainties due to
WIMP-nucleon currents. (Right) Integrated energy spectra of xenon for elastic and inelastic,
spin-dependent scattering. The inelastic contributions dominate
over the elastic ones for moderate energy thresholds.  A  WIMP mass of 100\,GeV, and a WIMP-nucleon cross section
10$^{-40}$\,cm$^2$ are assumed. Figure from \cite{Baudis:2013bba}. \label{fig:recoilspectra}}
\end{figure}

Spin-dependent inelastic scattering offers an additional and complementary detection channel, in particular in nuclei with low-lying excited states in the energy range $10-100$\,keV.  The experimental signature is a nuclear recoil together with the prompt de-excitation photon, boosting the region of interest to higher energies. Structure functions for inelastic WIMP-nucleus scattering are calculated in \cite{Baudis:2013bba}, showing that for momentum transfers around 150\,MeV, the inelastic channel is comparable to, or can even dominate the elastic one.  As an example, Figure~\ref{fig:recoilspectra} shows the differential recoil spectra for inelastic and elastic scattering off $^{129}$Xe and $^{131}$Xe (left), as well as  the integrated inelastic and elastic rates (right),  for 100\,GeV WIMP with a WIMP-nucleon cross section of 10$^{-40}$\,cm$^2$ \cite{Baudis:2013bba}.

\section{Backgrounds}
\label{sec:backgrounds}

A continuous challenge for direct dark matter search experiments is to minimize and accurately characterize the background noise. Some of the main background sources are the environmental radioactivity including airborne radon and its daughters, radio-impurities in the detector construction and shield material, neutrons from $(\alpha,n)$ and fission reactions, cosmic rays and their secondaries, and activation of detector materials during exposure at the Earth's surface \cite{Heusser:1995wd}.  Background sources intrinsic to the detector materials, muon-induced high-energy neutrons and ultimately neutrinos start to play an increasingly important role already for the current generation of detectors. 

In the past, a combination of low-Z and high-Z materials were employed to diminish the neutron and gamma fluxes coming from the laboratory walls and outer shield layers, and shield structures were kept under a N$_2$-atmosphere at slight overpressure to suppress the background from airborne radon decays and subsequent $^{210}$Pb plate-out.  Detectors recently constructed or planned for the future increasingly use large water shields, which passively reduce the environmental radioactivity and muon-induced neutrons, and act as an active muon veto. Typical underground gamma- and radiogenic neutron fluxes of 0.3\,cm$^{-2}$s$^{-1}$ and  9$\times$10$^{-7}$\,cm$^{-2}$s$^{-1}$, respectively \cite{Haffke:2011fp} are reduced by a factor of 10$^6$ after 3\,m and 1\,m of water shield \cite{Selvi:2011zz}. 

Critical are interaction from  $(\alpha,n)$- and fission neutrons from $^{238}$U and $^{232}$Th decays in detector components close to the dark matter target materials. The neutron energy spectra and yields are  calculated using the exact composition of these materials and the measured amount of  $^{238}$U and $^{232}$Th  in each component. Because often secular equilibrium in the primordial decay chains is lost in processed materials, the  $^{238}$U and $^{232}$Th activities are typically determined via mass spectrometry or neutron activation analysis, while the activities of the late part of these chains, $^{226}$Ra,  $^{228}$Ac, $^{228}$Th, are determined via gamma spectrometry using ultra-low background HPGe detectors \cite{Baudis:2011am}.  The neutrons are then transported using Monte Carlo simulations to evaluate the expected number of single-scatter nuclear recoils, which might be difficult to distinguish from a potential WIMP signal \cite{Mei:2008ir}. 

In liquified noble gas detectors, sources of radioactivity intrinsic to the WIMP target materials such as $^{39}$Ar, $^{85}$Kr and radon diffusion provide challenging backgrounds for current and possibly future detectors. For instance, impurity levels of $\sim$1\,ppt in natural krypton and $\sim$1\,$\mu$Bq/kg radon must be achieved by ton-scale experiments aiming to probe WIMP-nucleon cross sections down to 10$^{-47}$cm$^2$, with typical expected  background rates from external sources below 1 event per ton of target material and year.

\begin{figure}[h!]
\begin{minipage}[]{0.5\textwidth}
\begin{center}
\includegraphics*[width=1.\textwidth]{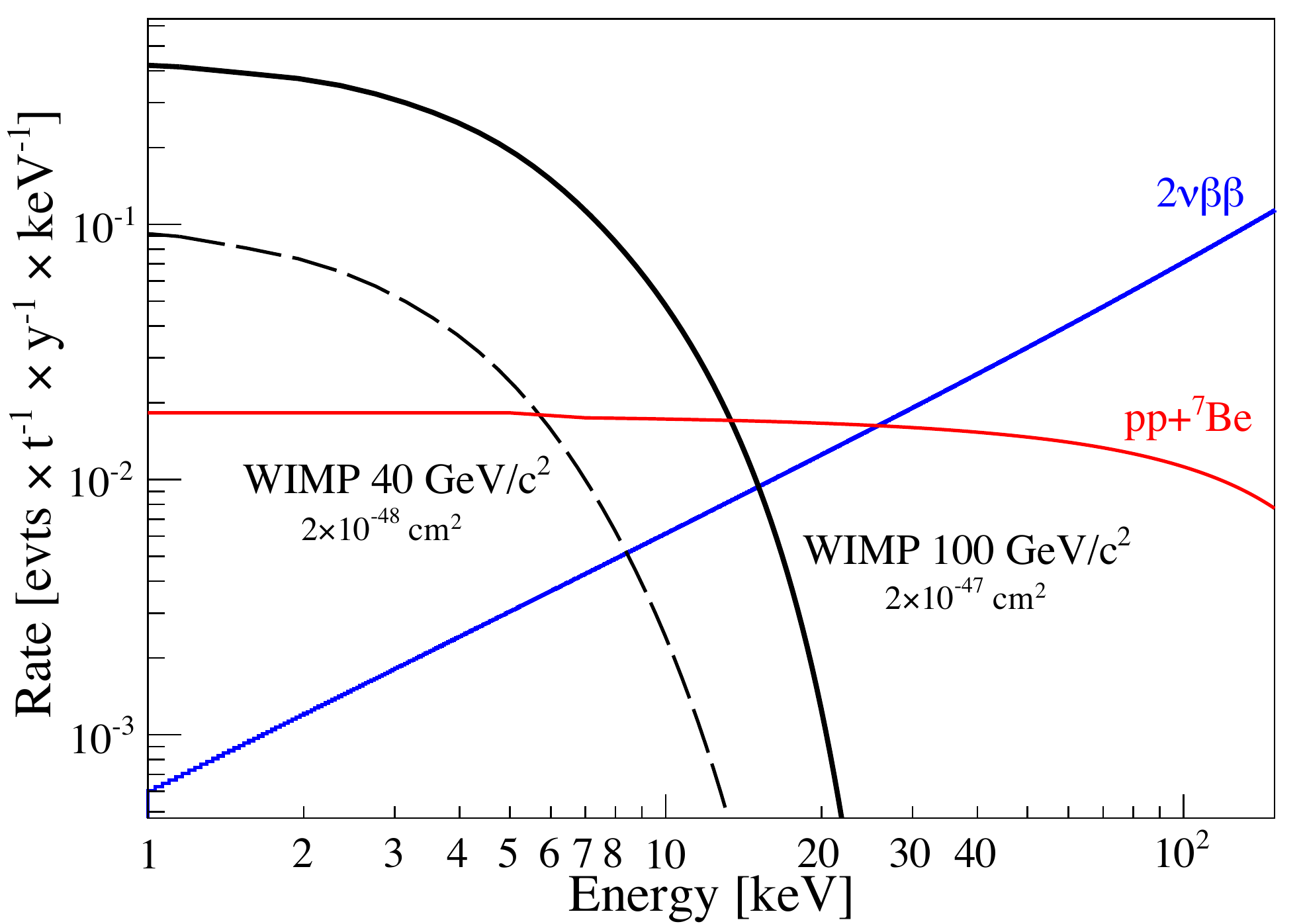}
\end{center}
\end{minipage}
\hfill
\begin{minipage}[]{0.5\textwidth}
\begin{center}
\includegraphics*[width=1.\textwidth]{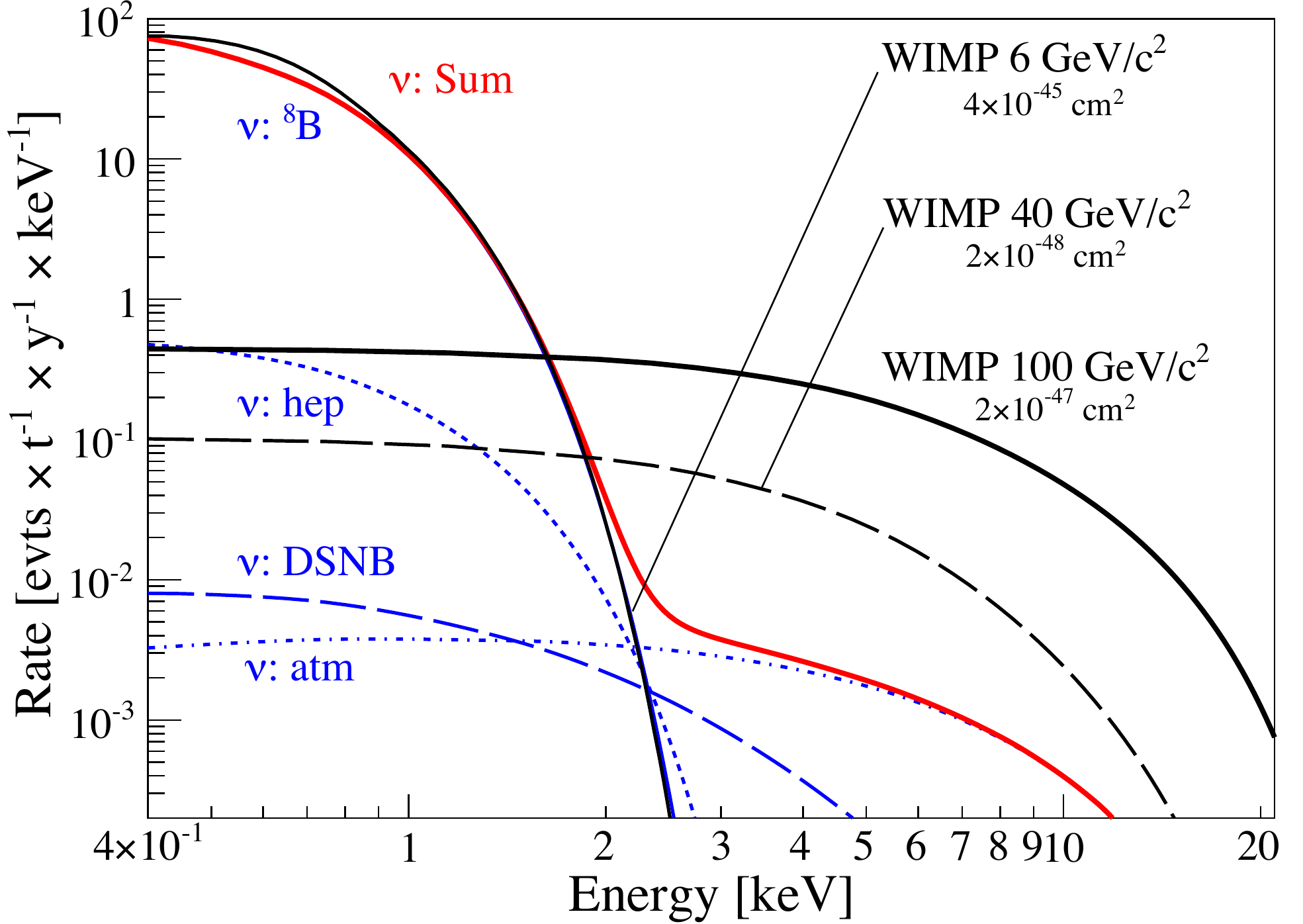}
\end{center}
\end{minipage}
{\caption{\small (Left)  Summed differential energy spectrum  for pp and  $^{7}$Be neutrinos (red) in a LXe detector. The electron recoil spectrum from the double beta decay of $^{136}$Xe (blue), as well as the expected nuclear recoil spectrum from WIMP scatters for a spin-independent WIMP-nucleon cross section of 2$\times$10$^{-47}$\,cm$^2$ (solid black) and 2$\times$10$^{-48}$\,cm$^2$ (dashed black) and  WIMP masses of 100\,GeV/c$^2$ and 40\,GeV/c$^2$ is also shown.  A 99.5\% discrimination of electronic recoils is assumed.  (Right): The  differential nuclear recoil spectrum from coherent scattering of neutrinos (red)  from the Sun, the diffuse supernova background (DSNB), and the atmosphere (atm), compared to the one from WIMPs  for various masses and cross sections (black). 
The coherent scattering rate will provide an irreducible background for low-mass WIMPs, limiting the cross 
section sensitivity to $\sim 4\times10^{-45}$\,cm$^2$ for WIMPs of 6\,GeV/$c^2$ mass, while WIMP masses 
above $\sim10$\,GeV/$c^2$ will be significantly less affected. For both plots, the nuclear recoil signals are converted 
to an electronic recoil scale (see \cite{Baudis:2013qla}) and a nuclear recoil acceptance of 50\% is assumed. Figure from~\cite{Baudis:2013qla}. \label{fig::wimp_neutrino}}}
\end{figure}

The ultimate background source might be provided by the irreducible neutrino flux. Solar pp-neutrinos will contribute to the electronic recoil background via neutrino-electron scattering at the level of $\sim$10-25\,events/(ton$\times$year) in the low-energy, dark matter signal region of a typical detector. Depending on the detector's  discrimination capabilities between electronic and nuclear recoils, solar pp-neutrinos may thus become a relevant background at cross sections $<$10$^{-48}$cm$^2$.  Neutrino-induced nuclear recoils from coherent neutrino-nucleus scatters can not be distinguished from a WIMP-induced signal. The $^{8}$B solar neutrinos yield up to 10$^3$\,events/(ton$\times$year) for heavy targets such as xenon \cite{Strigari:2009bq}, however most of these events are below the energy thresholds of current and likely also future noble liquid detectors. Nuclear recoils from atmospheric neutrinos and the diffuse supernovae neutrino background will yield event rates in the range 1- 5\,events/(100\,ton$\times$year), depending on the target material, and hence will dominate measured spectra at a WIMP-nucleon cross section below 10$^{-49}$cm$^2$  \cite{Strigari:2009bq,Gutlein:2010tq,Billard:2013qya,Anderson:2011bi,Baudis:2013qla}.  Figure \ref{fig::wimp_neutrino} shows the expected background spectrum from pp and $^7$Be solar neutrinos (left), as well as from coherent scattering of solar, atmospheric and diffuse supernova background neutrinos (right) in a detector using xenon as target material. Also shown are predicted WIMP-induced differential nuclear recoil spectra for various WIMP masses and cross sections. The underlying assumptions are detailed in the figure caption, and in \cite{Baudis:2013qla}.

\section{Liquefied noble gases as WIMP targets}

From all noble elements, only argon and xenon are currently used as targets for dark matter detection. Neon has been suggested as a medium for low-energy neutrino detection, and  could potentially be employed in the search for WIMPs \cite{Lippincott:2011zr}. In their liquid phase, noble elements are excellent media for building  large, homogeneous, compact and self-shielding detectors.  Liquid xenon (LXe) and liquid argon (LAr) are excellent scintillators and good ionizers in response to the passage of radiation. The simultaneous detection of ionization and scintillation signals allows to identify the primary particle interacting in the liquid, for the ratio of the two observables depends on $dE/dx$.  In addition, the 3D position of an interaction can be determined with sub-mm (in the z-coordinate) to mm (in the x-y-coordinate) precisions in a time projection chamber (TPC).   These features, together with the relative ease of scale-up to large detector masses, have contributed to make LXe and LAr powerful targets for WIMP searches  \cite{Aprile:1900zz,Chepel:2012sj}. 

Table \ref{Table1} summarizes the physical properties of argon and xenon. Their atomic number, density, boiling point temperature, dielectric constant, abundance in the atmosphere and intrinsic radioactive isotopes determine practical aspects of a dark matter detector. Xenon is heavier and its high liquid density helps to design a compact detector geometry with efficient self-shielding. Its fraction in the atmosphere is very low, making it more expensive than natural argon. Xenon does not contain radioactive isotopes, apart from the double beta emitter $^{136}$Xe, with a half-life recently measured to be (2.165$\pm$0.016$\pm$0.059)$\times$10$^{21}$\,yr \cite{Albert:2013gpz}.  

\begin{table}[h!]
\begin{center}
\label{Table1}
\begin{tabular}{|l|c|c|}
\hline
{Property [unit]} & {Xe} & {Ar}  \\ 
\hline
Atomic number & 54  &  18  \\
Mean atomic weight & 131.3 & 40.0 \\
Boiling point $T_b$ at 1 atm [K] & 165.0 & 87.3  \\
Melting point $T_m$ at 1 atm [K]& 161.4 & 83.8  \\
Gas density at 1 atm \& 298 K [g/l] & 5.40 & 1.63 \\
Gas density at 1 atm \& $T_b$ [g/l] & 9.99 & 5.77  \\
Liquid density at $T_b$ [g/cm$^3$] & 2.94 & 1.40  \\
Volume ratio & 526 & 795 \\ 
Dielectric constant of liquid & 1.95 & 1.51  \\
Volume fraction in Earth's atmosphere [ppm] & 0.09 & 9340 \\
Radioactive isotopes & $^{136}$Xe, T$_{1/2}$ = 2.16$\times$10$^{21}$yr & $^{39}$Ar, T$_{1/2}$ = 269\,yr\\
\hline
\end{tabular}
\caption{Physical properties, volume fraction in the atmosphere and radioactive isotopes of the noble elements xenon and argon.}
\end{center}
\end{table}

Atmospheric argon contains the radioactive $^{39}$Ar, with a measured ratio of $^{39}$Ar to $^{40}$Ar of 8.1$\times$10$^{-16}$\,g/g, resulting in a specific activity of  $^{39}$Ar  of 1\,Bq/kg \cite{Benetti:2006az}. Its production by cosmic ray induced reactions is dominated by the process  $^{40}$Ar(n,2n)$^{39}$Ar, which has an energy threshold around 1\,MeV. This increases the effective costs of argon for large dark matter detectors, since depletion in $^{39}$Ar by isotopic separation or by extraction from underground wells is necessary.  Depletion factors of larger than 100 have been achieved by using argon extracted from underground gas wells \cite{Back:2012pg}. An anthropogenic radioactive isotope present in noble liquids extracted from air is $^{85}$Kr. Currently achieved $^{nat}$Kr-levels after purification using krypton-distillation columns are (1.0$\pm$0.2)\,ppt  \cite{Lindemann:2013kna} (with a detection limit by gas chromatography and mass spectrometry of 0.008\,ppt), (3.5$\pm$1.0)\,ppt \cite{Akerib:2013tjd} and $<$2.7\,ppt \cite{Abe:2013tc}.  Levels of  $\sim$ 0.1\,ppt will be necessary for an ultimate noble liquid dark matter detector ~\cite{Baudis:2013qla}.

The energy loss of an incident particle in noble liquids is shared between ionization, excitation  and sub-excitation electrons liberated in the ionization process. The average energy loss in ionization is slightly larger than the ionization potential or the gap energy because it includes multiple ionization processes.   Scintillation from noble liquids arises from excited atoms R* and from ions R$^+$, both being produced by ionizing radiation (see \cite{Aprile:1900zz,Chepel:2012sj} and references therein):\\

\begin{enumerate}
\item 
\[{\mathrm R}^*+{\mathrm R}+{\mathrm R}\rightarrow {\mathrm R}^{*2}+{\mathrm R}\]
\[{\mathrm R}^{*2}\rightarrow 2{\mathrm R}+{\mathrm h\nu}\] 
\item 
\[{\mathrm R}^{+}+{\mathrm R} \rightarrow 2{\mathrm R}^{+2}\]
\[{\mathrm R}^{+2}+e^-\rightarrow {\mathrm R}^{**}+{\mathrm R}\]
\[{\mathrm R}^{**}\rightarrow {\mathrm R}^*+{\mathrm {heat}}\]
\[{\mathrm R}^*+{\mathrm R}+{\mathrm R}\rightarrow {\mathrm R}^{*2}+{\mathrm R}\]
\[{\mathrm R}^{*2}\rightarrow 2{\mathrm R}+{\mathrm h\nu}\]
\end{enumerate}

\noindent
where h$\nu$ denotes the emitted vacuum-ultraviolet (VUV) photons with wavelength of 178\,nm and 128\,nm for LXe and LAr, respectively.  R**$\rightarrow$R* + heat corresponds to a non-radiative transition. In both processes, the excited dimer R*$^2$, at its lowest excited level, is de-excited to the dissociative ground state by the emission of a single UV photon. This comes from the large energy gap between the lowest excitation and the ground level, forbidding other decay channels such as non-radiative transitions.  The scintillation light from pure liquid argon and xenon has two decay components due to de-excitation of singlet and triplet states of the excited dimer ${\mathrm R}^{*2}\rightarrow 2{\mathrm R}+{\mathrm h\nu}$.  The shorter decay shape is produced by the de-excitation of singlet states and the longer one by the de-excitation of triplet states. The differences of pulse shape between different type of particle interactions can be used to  discriminate electronic from nuclear recoils, using the pulse shape discrimination (PSD) technique. It is particularly effective for liquid argon, given the large separation in time of the two decay components \cite{Hitachi:1983zz,Lippincott:2008ad}.  

%\newpage

In a given detection medium, the deposited energy for electronic and nuclear recoils can be expressed as (see \cite{Sorensen:2011bd} and references therein):

\begin{equation}
E_{er} = w \left(n_{\gamma} + n_{e^-}\right)
\end{equation}

\begin{equation}
E_{nr} = w \left(n_{\gamma} + n_{e^-}\right) \frac{1}{\mathcal{L}}
\end{equation}
where $w$ is the average energy required for the production of a single photon or an electron, $n_{\gamma}$ and $n_{e^-}$ are the number of produced photons and electrons, respectively, and ${\mathcal{L}}$ is the so-called Lindhard-factor. It quantifies the quenching of the observed signal for nuclear recoils, compared to the one for  electronic recoils of the same energy. The $w$-value is  $\sim 14$\,eV and $\sim24$\,eV  in LXe and LAr, respectively.  The connection to actual signals observed in a dark matter detector, namely the prompt (S1) and proportional  (S2) scintillations (the latter only in the case of dual-phase time projection chambers), is the following:

\begin{equation}
n_{\gamma} = \frac{S1}{\epsilon_1}, \,\,\,\,\, n_{e^-} = \frac{S2}{\epsilon_2},
\end{equation}
where ${\epsilon_1}$ is the photon detection efficiency (in phe/$\gamma$) and ${\epsilon_2}$ is the average number of photoelectrons (phe) per extracted electron into the vapour phase (in phe/e$^-$), where 100\% electron extraction efficiency is assumed.

The average number of photoelectrons per extracted electron can be measured {\sl in situ}; it is for instance  $\sim$20\,phe/$e^-$ in the XENON100  \cite{Aprile:2013blg}, and $\sim$25\,phe/$e^-$ in the LUX \cite{Szydagis:2014xog} xenon detectors. The photon detection efficiency is more difficult to determine directly, it is typically $\sim$0.10-0.15\,phe/$\gamma$. Experiments  thus usually report the light yield $L_y$ (in phe/keV) for electronic recoils in a detector, which is measured at several energies using mono-energetic electron and gamma lines from a variety of sources. It includes the photon detection efficiency, geometrical effects and the quantum efficiency of the light sensors.  In the past, the "standard candle" was taken as the 122\,keV gamma line of $^{57}$Co,  most recently it shifted to the lower energy, 9.4\,keV or the 32.1\,keV, lines from $^{83m}$Kr decays, for $^{83m}$Kr can be used as an internal calibration source in both argon and xenon detectors \cite{Manalaysay:2009yq,Kastens:2009pa}.  While for the 9.4\,keV emission which follows the 31.2\,keV line with a half-life of 154.4\,ns,  a time dependence of the scintillation signal is observed, the 32.1\,keV transition shows no time dependance, and can safely be used as a "standard candle"  \cite{Baudis:2013cca}. The Noble Element Simulation Technique (NEST) considers all existing measurements of the light and charge yields for electronic and nuclear recoils (mostly in xenon, but also in argon) and it is used to simulate a given process in a single phase detector, or in a time projection chamber \cite{Szydagis:2013sih,Szydagis:2011tk}. 

In dark matter detectors using noble liquids, the energy of the nuclear recoil was traditionally established based on the observation of the primary scintillation signal  alone.  The scintillation light yield of nuclear recoils is different from the one produced by electron recoils of the same energy and the ratio at zero drift field, termed relative scintillation efficiency, $\mathcal{L}_{\rm{eff}}$,  is defined as follows:
\begin{equation}
\mathcal{L}_{\rm{eff}} (E_{nr})= \frac{S1}{L_y}\cdot \frac{1}{E_{nr}},
\end{equation}
where  $L_y$ is the light yield of electronic recoils at a given energy, as mentioned above.  At non-zero drift field, a correction factor $S_{er}/S_{nr}$ must be applied, where $S_{er}$, $S_{nr}$ parametrize the suppression of the electronic and nuclear recoil signals due to the presence of the field.

For LXe, the relative scintillation efficiency of nuclear recoils has been measured down to 3\,keV nuclear recoil energy (keV$_{nr}$) \cite{Plante:2011hw} at zero drift field.  $\mathcal{L}_{\rm{eff}}$  is $\sim$9\%,  for nuclear recoils of 3\,keV$_{nr}$  and $\sim$14\% at 15\,keV$_{nr}$. For higher energy nuclear recoils, the value is on average $\sim$19\%. These measurements agree with indirect methods based on the data/MC comparison of nuclear recoil spectra in a given detector \cite{Aprile:2013teh}, and with a recent, in situ measurement by LUX, using a mono-energetic neutron beam and double-scatters to tag nuclear recoils \cite{lux_ucla2014}.  The quantities $S_{er}$ and $S_{nr}$ depend on the applied electric field, but seem to be independent of energy. However more accurate, direct measurements are needed.  In LAr, $\mathcal{L}_{\rm{eff}}$  at zero field has been measured down to 10\,keV$_{nr}$ and 16\,keV$_{nr}$, respectively \cite{Gastler:2010sc,Regenfus:2012kh}. An average value of 25\% and 29\% above an energy of  20\,keV$_{nr}$ and  16\,keV$_{nr}$, is observed, with a turn up at lower recoil energies.  A field dependance of up to 32\% of the scintillation for nuclear recoils of energies between 10\,keV$_{nr}$ and 50\,keV$_{nr}$ has  been observed \cite{Alexander:2013aaq}, which increases with decreasing energy. Recently $\mathcal{L}_{\rm{eff}}$ in LAr was measured down to about 10\,keV$_{nr}$, and no turn up at energies below 20\,keV$_{nr}$ is seen \cite{Lippincot_lhcforum_2014}.

Nuclear recoils generated by WIMPs and fast, MeV neutrons have denser tracks, and  thus greater electron-ion recombination than electron recoils. The collection of ionization electrons thus becomes more difficult for nuclear than for electron recoils. The ionization yield $Q_y$ of nuclear recoils is defined as the number of observed electrons per unit recoil energy ($\rm e^-/keV_{nr}$): 
\begin{equation}
Q_y(E_{nr}) = \frac{S2}{\epsilon_2}\cdot \frac{1}{E_{nr}}.
\end{equation}

In xenon, $Q_y$ was measured for the first time by \cite{Aprile:2006kx}, as a function of external electric field and recoil energy. An increase in  charge yield for low energy recoils was observed, explained by the  lower recombination rate expected by the drop in electronic stopping power at low energies. This behaviour, although theoretically not fully understood, allows  the observation of Xe  nuclear recoils down to a few keV$_{nr}$, improving the event rate and the sensitivity for WIMP detection. Indirect measurements, obtained from a comparison of data and Monte Carlo simulations in a given detector down to 3\,keV$_{nr}$ \cite{Aprile:2013teh}, agree with previous data \cite{Manzur:2009hp} and with the model by \cite{Bezrukov:2010qa}. The ionization yield  of nuclear recoils in LAr has been first measured by \cite{Grandi:2005dm,Benetti:2007cd} and recently by the SCENE collaboration \cite{Lippincot_lhcforum_2014,Cao:2014gns}.  It also shows significantly less yield than the one from electronic recoils and an increase in the yield for lower recoil energies. In addition, an increase in the charge yield is observed for larger drift fields \cite{Lippincot_lhcforum_2014,Cao:2014gns}.  

For electronic recoils, the light yield in LXe  was measured down to 1.5\,keV recoils, with \cite{Baudis:2013cca} and without \cite{Aprile:2012an} an applied drift field. At zero field, a reduced scintillation yield is observed below 10\,keV, dropping to $\sim$40\% of its value at higher energies. With an applied field of 450 V/cm, a reduction of the scintillation output to about 75\% relative to the value at zero field is observed. These measurement are particularly relevant for axion and axion-like-particle searches with liquid xenon detectors \cite{Abe:2012ut,Aprile:2014eoa}, as well as for searches of other dark matter candidates that might interact with electrons, and of neutrino-electron scattering \cite{Baudis:2013qla}.

\subsection{The single and dual-phase detector technology}
\label{sec:singlephase}

In single-phase liquid argon or xenon dark matter detectors, the experimental strategy is to use the self-shielding of the liquid in order to achieve an effectively background-free inner volume, coupled to pulse shape discrimination in the case of argon. The self-shielding works better for the gamma-, than for the neutron background, given the differences in their mean free paths. The position of an interaction can be reconstructed using the  observed light pattern in the photodetectors, which are usually mounted on the inner detector surface. The position resolution will depend on the size of the  photodetectors, as well as on the overall light yield.  The main advantage of these detectors is their simplicity compared to time projection chambers. Since no electrons are drifted, the need for a high-voltage and a cathode-anode-grid system falls away.   It goes at the cost of a large amount of expensive (in the case of xenon, and of argon depleted in $^{39}$Ar) cryogenic liquid which is used as shield.  

A disadvantage, in particular for xenon detectors, rests in the fact that less information per event becomes available. In liquid xenon, there is no well-defined separation between electron and nuclear recoils based on the pulse shape of events, since the mean lifetime of the singlet and triplet states are not well separated in time (3\,ns versus 27\,ns). These detectors thus rely on the absolute suppression  of both external and internal electronic recoil sources.   In LAr, the decay times  of the singlet and triplet state differ by two orders of magnitude, being $\tau_s$ = 7\,ns and $\tau_t$ = 1.6\,$\mu$s, respectively.  The ratio between the emission intensities of the singlet/triplet states depends on the deposited energy per unit path length, $dE/dx$. It has been measured to be about 0.3 for gammas, in the range of 1.3-3 for alphas and about 3 for argon ion recoils \cite{Hitachi:1983zz, Lippincott:2008ad,Peiffer:2008zz}. The difference in the pulse shapes is employed to distinguish between electronic and nuclear recoils, where a discrimination power of $\sim$10$^{6}$ has been achieved for nuclear recoil energies above $\sim$50\,keV \cite{Lippincott:2008ad}. Because of the shorter wavelength of the scintillation light in LAr (128\,nm), the inner detector surfaces and the PMT windows must be coated with a wavelength  shifter, such as for instance tetra phenyl butadiene (TPB), which shifts the wavelength of the light to about 442\,nm \cite{Amsler:2007gs}.  One must ensure an optimal efficiency of the wavelength shifter, and its long-term stability in time in a cold, LAr environment.

Two-phase detectors rely on the time projection chamber technique. An interaction within the so-called active volume of the detector will create ionization electrons and prompt scintillation photons, and  both signals are detected. The electrons drift in the pure liquid under the influence of an external electric field, are then accelerated by a stronger field and extracted into the vapour phase above the liquid, where they generate proportional scintillation, or electroluminiscence.  Two arrays of photosensors, one in the liquid and one in the gas, detect the prompt scintillation (S1) and the delayed, proportional scintillation signal (S2). The array immersed in the liquid collects the majority of the prompt signal, which is totally reflected at the liquid-gas interface. The ratio of the two signals is different for nuclear recoils  created by {\small WIMP} or neutron interactions, and electronic recoils produced by $\beta$ and $\gamma$-rays, providing the basis for background discrimination. 
Since electron diffusion in the ultra-pure liquid is small, the proportional scintillation photons carry the $x-y$ information of the interaction site. With the $z-$information from the drift time measurement, the {\small TPC} yields  a three-dimensional event localization, enabling to reject the majority of the background via fiducial volume cuts.

\subsection{Current xenon experiments}

The most stringent limits  for spin-independent WIMP-nucleon couplings to date come from the dual-phase LXe TPCs ZEPLIN-III \cite{Akimov:2011tj,Akimov:2010zz}, XENON10 \cite{Angle:2007uj,Aprile:2010bt}, XENON100 \cite{Aprile:2011hi,Aprile:2011dd,Aprile:2012nq}  and LUX \cite{Akerib:2013tjd}, where XENON100 and LUX are currently running experiments.  XENON100,  a 161\,kg  LXe TPC, is operated  in the interferometer tunnel at LNGS in Italy. It employs two arrays of low-radioactivity, UV-sensitive photomultipliers (PMTs) to detect the prompt and proportional light signals induced by particles interacting in the xenon volume, containing 62\,kg of ultra-pure liquid xenon. The remaining 99\,kg of LXe act as an active veto shield against background events.  The instrument is described in \cite{Aprile:2011dd}, the analysis procedure is detailed in \cite{Aprile:2012vw}. Using 13 months of data taken during 2011 and 2012, with an exposure of 224.6\,live days $\times$ 34\,kg, XENON100 had reached its initial aim to probe spin-independent WIMP-nucleon cross sections down to 2$\times$10$^{-45}$cm$^2$ at a 55\,GeV WIMP mass \cite{Aprile:2012nq}. New limits on WIMP-nucleon  spin-dependent couplings were also derived \cite{Aprile:2013doa}, yielding the world's best sensitivity on WIMP-neutron couplings. After a further distillation run to reduce the Kr concentration to (0.95$\pm$0.16)\,ppt of $^{nat}$Kr in LXe  in the latest XENON100 run \cite{Lindemann:2013kna}, a new  dark matter run started in 2013, and lasted for 150 days. The blind analysis of this data is in progress.

The Large Underground Xenon (LUX) experiment is a 370\,kg (250\,kg) total (active) mass LXe TPC in a water Cherenkov shield at the Sanford Underground Research Facility (SURF) in Lead, South Dakota, USA \cite{Akerib:2012ys}. The TPC is 47\,cm in diameter and 48\,cm in height, the S1 and S2 signals are detected via two arrays of 61 PMTs, one in the liquid and one in the gas region. A first science run lasted  from April to August 2013, for 85.3 live-days. A non-blind analysis was performed using a fiducial volume of 118\,kg and a software imposed energy threshold of 3\,keV$_{nr}$.  The reached 90\% upper C.L. cross section for spin-independent WIMP-nucleon couplings has a minimum of  $7.6 \times10^{-46}$\,cm$^2$ at a WIMP mass of 33\,GeV/c$^2$, thus confirming and improving upon the XENON100 results. LUX plans to continue operations at SURF in 2014 and 2015, and to reach its sensitivity goal with a blinded 300 live-day WIMP search \cite{Akerib:2013tjd}.

The XMASS collaboration operates a large, single-phase LXe detector with 835\,kg in the sensitive region, at the Kamioka Observatory in the Kamioka mine in Japan \cite{Moriyama:2011zz}.  The inner detector is equipped with 642  hexagonal PMTs in an almost spherical shape, contained in a copper vessel filled with LXe. The outer detector, used as an active shield for cosmic muons, is a cylindrical water tank, 10\,m in height and 10\,m in diameter, equipped with 70 20-inch PMTs.  XMASS recently published a result for low-mass WIMPs~\cite{Abe:2012az} as well as limits on inelastic WIMP-$^{129}$Xe scatters \cite{Uchida:2014cnn}.  The collaboration is currently operating a refurbished detector with reduced background levels, with new results expected for 2014. 

PandaX is a dual-phase xenon TPC at the Jinping laboratory in Sichuan, China, operated in a conventional low-background shield \cite{Cao:2014jsa}. In its first stage, it plans to increase the fiducial mass from 25\,kg to 300\,kg of xenon, with a total target mass of about 500\,kg. The active xenon region is viewed by 150 1-inch R8520 PMTs on the top, and 40 3-inch R11410-10 PMTs on the bottom array.  PandaX completed two engineering runs in 2013, and is currently preparing the first physics run with 450\,kg of xenon in the system \cite{pandax_ucla2014}.

XENON1T, the next step of the XENON program, is a dual-phase xenon TPC with total (active) mass of 3.3\,t ($\sim$2\,t).  The active LXe volume will be viewed by 248 Hamamatsu R11410-21 3-inch PMTs, arranged in two arrays. The PMTs, with an average quantum efficiency of 36\% at 175\,nm, were extensively tested for their long-term stability in liquid xenon \cite{Baudis:2013xva}.  The underground construction of XENON1T started in Hall~B of LNGS in the fall of 2013, completion and commissioning are expected for early and mid 2015, respectively.  With an aimed background which is two orders of magnitude below the one from XENON100, a sensitivity to spin-independent cross sections of $2\times10^{-47}$\,cm$^2$ is expected after two years of continuous operation underground.

\subsection{Current argon experiments}

DarkSide-50~\cite{Bossa:2014cfa} is a dual-phase argon detector with 50\,kg (33\,kg) active (fiducial) mass, viewed by 38 Hamamatsu R11065 3-inch PMTs. The TPC is surrounded by a 30\,t borated-liquid scintillator neutron veto and  both detectors are deployed in the Borexino Counting Test Facility (CTF), a 1\,kton water Cherenkov shield  in Hall C of  LNGS. The collaboration has recently presented first results from a run with natural argon, demonstrating a light yield of 8\,phe/keV at 41.5\,keV and zero field, and an electron-lifetime $>$5\,ms for a maximum drift time of 370\,$\mu$s.  A background free exposure of 280\,kg\,day allowed to prove the efficiency of the PSD, predicting no background in 2.6\,y of a run with argon depleted in the radioactive isotope $^{39}$Ar. The projected sensitivity using the measured PSD performance, an energy threshold of 35\,keV$_{nr}$ and a data taking period of 3 years is $\sim$10$^{-45}$cm$^2$ for a 100\,GeV WIMP \cite{grandi_ucla2014}.  

The  ArDM experiment operates a ton-scale dual-phase Ar TPC at the Canfrac Underground Laboratory in Spain~\cite{Marchionni:2010fi}. The active detector volume, containing 850\,kg of LAr, is viewed by two arrays of 12 8-inch PMTs arranged in a honeycomb pattern. TPB is deposited on the PMT windows  and on a Tetratex reflector installed on the field shaping rings.  The detector and its infrastructure were installed underground successfully, and first data acquired with the detector filled with pure Ar gas at room temperature has allowed the assessment of the light detection system's performance \cite{Badertscher:2013ygt}. Commissioning with liquid argon is expected for mid 2014, and first results are expected for 2015.

The MiniCLEAN~\cite{Rielage:2014pfm} and DEAP-3600 \cite{Boulay:2012hq} experiments are single-phase LAr detectors detectors operated at SNOLab in Sudbury, Canada. MiniCLEAN  will operate 500\,kg of LAr in an inner stainless steel vessel of about 64-inch in diameter, contained in a vacuum cryostat. The assembly is located inside a large water shield operated as muon veto. The liquid argon is viewed by 92 optical modules, containing each one 8-inch Hamamatsu R5912-02Mod PMT, coupled to a light guide and an acrylic plug at the end. The plug front surfaces, coated with TPB,  form the 92-sided sphere that contains the cryogenic volume.  The MiniCLEAN detector is nearing completion underground, cooling with cold gas and liquidation of argon are scheduled to start in summer 2014 ~\cite{Rielage:2014pfm}.   The DEAP-3600 detector will operate  3600\,kg LAr (with 1000\,kg fiducial mass), in a large acrylic vessel with an inner radius of 85\,cm. The LAr is viewed by 255 8-inch PMTs through 50\,cm light guides. These provide thermal insulation between the cryogenic acrylic vessel and the PMTs, as well as a shield against neutrons. The inner detector is housed in a spherical stainless steel vessel immersed in an 8\,m diameter water tank \cite{Boulay:2012hq}. The detector is under construction at SNOLab and first results are expected within one year.  The sensitivity aim with argon depleted in $^{39}$Ar is $<$$1\times10^{-46}$\,cm$^2$ at a WIMP mass of 100\,GeV. 

\subsection{Future projects and complementarity}

\begin{figure}[h!]
\begin{minipage}[]{0.6\textwidth}
\begin{center}
\includegraphics*[width=1.00\textwidth]{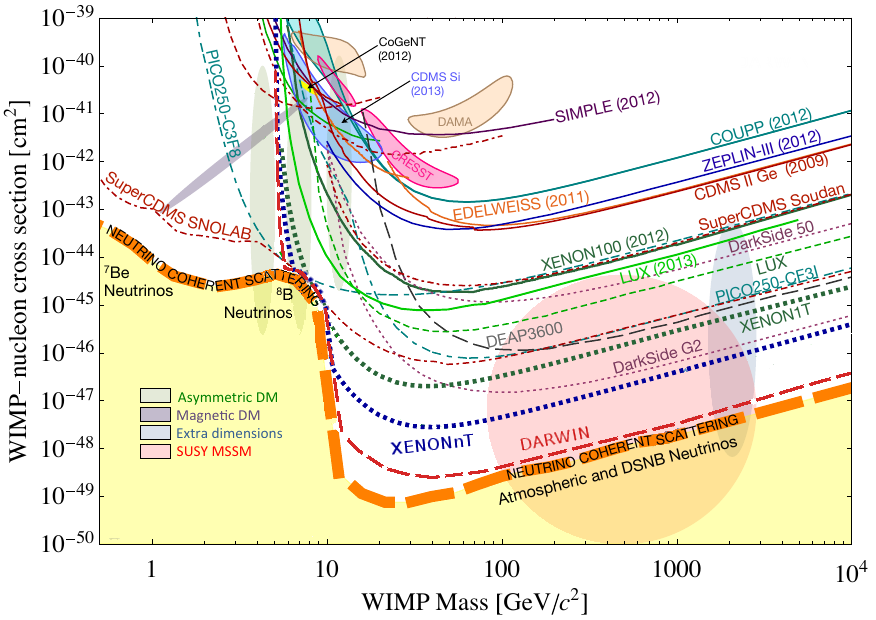}
\end{center}
\end{minipage}
\hfill
\begin{minipage}[]{0.4\textwidth}
\vspace{-0.8cm}
{\caption{\small Summary for spin-independent WIMP-nucleon scattering results. Existing limits from the noble gas dark matter experiments  ZEPLIN-III~\cite{Akimov:2011tj}, XENON10~\cite{Angle:2007uj}, XENON100~\cite{Aprile:2012nq}, and LUX~\cite{Akerib:2013tjd}, along with projections for  DarkSide-50 \cite{Bossa:2014cfa}, LUX \cite{Akerib:2013tjd}, DEAP3600 \cite{Boulay:2012hq}, XENON1T, DarkSide G2, XENONnT (similar sensitivity as the LZ project \cite{Malling:2011va}, see text) and DARWIN \cite{Baudis:2012bc} are shown.  DARWIN is designed to probe the entire parameter region for WIMP masses above $\sim$6\,GeV/c$^2$, until the neutrino background (yellow region) will start to dominate the recoil spectrum. Experiments based on the mK cryogenic technique such as SuperCDMS \cite{Agnese:2014aze} and EURECA \cite{Kraus:2011zz} have access to lower  WIMP masses. Figure adapted from~\cite{Cushman:2013zza}. \label{fig::limits}}}
\end{minipage}
\end{figure}

Existing results and projected sensitivities for the spin-independent WIMP-nucleon interactions as a function of the WIMP mass are summarized in Figure~\ref{fig::limits}, adapted from \cite{Cushman:2013zza}. In spite of observed anomalies in a handful of experiments, that could be interpreted as due to WIMPs, albeit not consistently, we have no convincing evidence of a direct detection signal induced by galactic dark matter. Considering LUX's lack of a signal in 85.3 live-days$\times$118\,kg of liquid xenon target, excluding $\sim$33\,GeV WIMPs with interaction strengths above  7.6$\times$10$^{-46}$cm$^2$, it becomes clear that, at the minimum, ton-scale experiments are required for a discovery above the 5-sigma confidence level (unless the WIMP is lighter than $\sim$10\,GeV, where larger cross sections are feasible). Several large-scale direct detection experiments are in their planning phase and will start science runs within this decade.

The next phase in the LUX program, LUX-ZEPLIN (LZ),  foresees a 7\,t LXe detector in the same SURF infrastructure, with an additional scintillator veto to suppress the neutron background. Construction is expected to start in 2014, and operation in 2016, with the goal of reaching  a sensitivity of 2$\times$10$^{-48}$cm$^2$ after three years of data taking \cite{Malling:2011va}. The upgrade of XENON1T, XENONnT,  is to increase the sensitivity by another order of magnitude, thus also reaching 2$\times$10$^{-48}$cm$^2$. While much of the XENON1T infrastructure will be reused, the inner detector will be designed and constructed once XENON1T is taking science data, with planned operation between 2018-2021. The XMASS collaboration plans a 5\,t (1\,t fiducial) single-phase detector after its current phase, with greatly reduced backgrounds and an aimed  sensitivity of $\sim$10$^{-46}$cm$^2$.  In its second stage, PandaX will operate a total of 1.5\,t LXe as WIMP target, with $\sim$1\,t xenon in the fiducial volume.  All sub-systems of the existing experiment,  with the exception of the central TPC, are designed to accommodate the larger target mass \cite{pandax_ucla2014}. The DarkSide collaboration plans a 5\,t LAr dual-phase detector, with 3.3\,t as active target mass, in the existing neutron and muon veto at LNGS. The aimed sensitivity is 10$^{-47}$cm$^2$ \cite{Alexander:2013hia}.
 
{DARk matter WImp search with Noble liquids} (DARWIN) is an initiative to build an ultimate, multi-ton dark matter detector at LNGS \cite{Baudis:2010ch, Baudis:2012bc}. Its primary goal is to probe the spin-independent WIMP-nucleon cross section down to the 10$^{-49}$\,cm$^2$ region for $\sim$50\,GeV/c$^2$ WIMPs, as shown in  Figure \ref{fig::limits}. It would thus explore the experimentally accessible parameter space, which will be finally limited by irreducible neutrino backgrounds. Should WIMPs be discovered by an existing or near-future experiment, DARWIN will measure WIMP-induced nuclear recoil spectra with high-statistics, constraining the mass and the scattering cross section of the dark matter particle~\cite{Pato:2010zk,Newstead:2013pea}. Other physics goals of DARWIN are the first real-time detection of solar pp-neutrinos with high statistics and the search for the neutrinoless double beta decay~\cite{Baudis:2013qla}. The latter would establish whether the neutrino is its own anti-particle, and can be detected via $^{136}$Xe, which has a natural abundance of 8.9\% in xenon.

\begin{figure}[h!]
\begin{minipage}[]{0.5\textwidth}
\begin{center}
\includegraphics[width=0.8\textwidth]{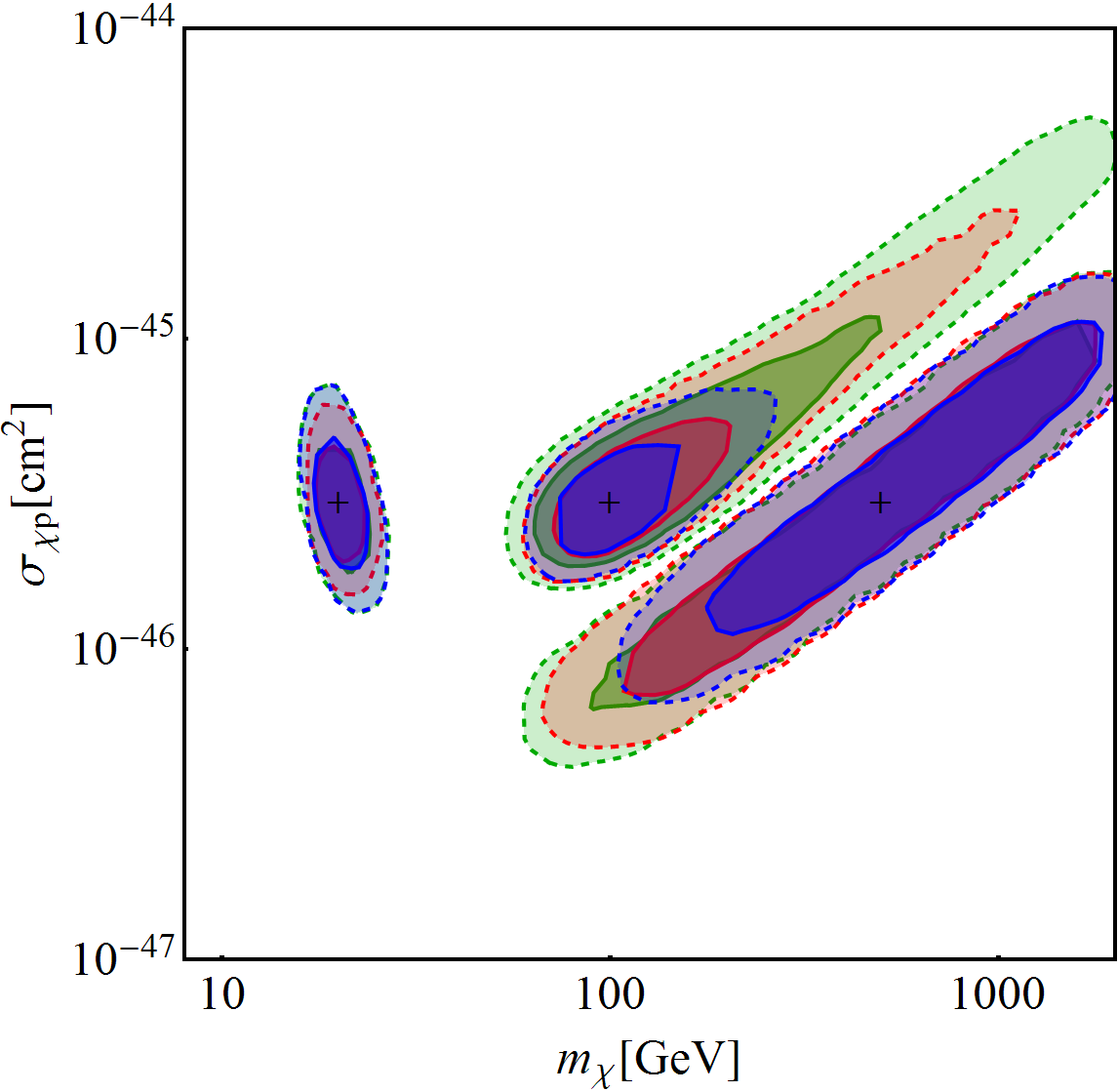}
\end{center}
\end{minipage}
\hfill
\begin{minipage}[]{0.5\textwidth}
\begin{center}
\includegraphics[width=0.8\textwidth]{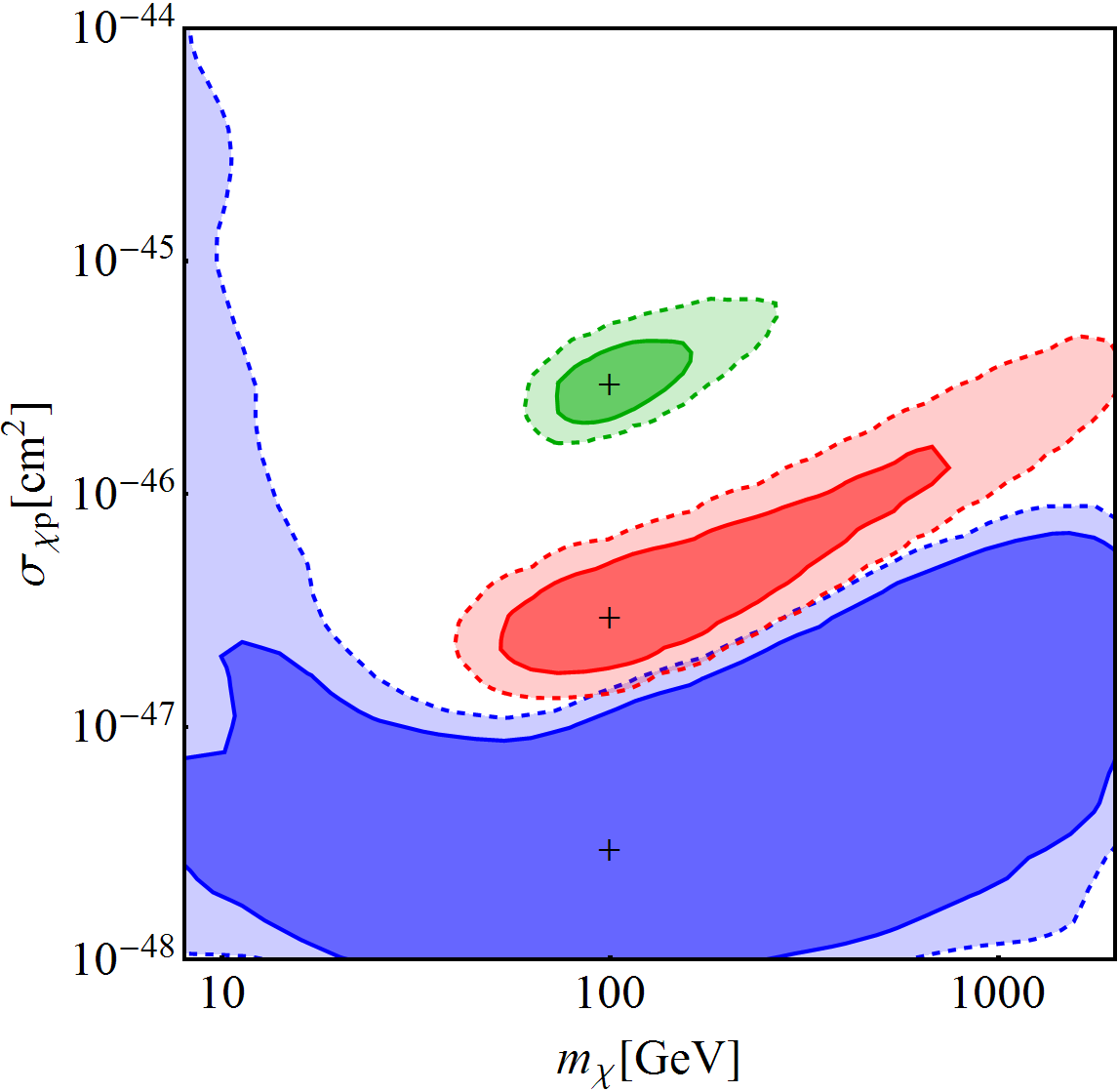}
\end{center}
\end{minipage}
{\caption{\small The 1-$\sigma$ and 2-$\sigma$ credible regions of the marginal posterior probabilities for simulations of WIMPs of various masses and cross sections demonstrate how well the WIMP parameters can be reconstructed for different exposures of a LXe or LAr detector. The `+' indicates the simulated model. (Left) Exposures of 10\,t$\times$yr xenon (green), 20\,t$\times$yr LXe (red), and 10\,t$\times$yr LXe plus 20\,t$\times$yr LAr (blue). (Right) 10\,t$\times$yr LXe plus 20\,t$\times$yr LAr exposure for cross sections of $3\times10^{-46}$\,cm$^2$ (green), \mbox{$3\times10^{-47}$\,cm$^2$} (red) and $3\times10^{-48}$\,cm$^2$ (blue).  Figure from \cite{Newstead:2013pea}.}}
\label{fig::sens}
\end{figure}

As soon as direct evidence for a dark matter signal has been established with so-called discovery experiments, the efforts will shift towards measuring the mass and cross section of the dark matter particle. This information is complementary to the one provided by indirect detection experiments such as IceCube, AMS and CTA, and by dark matter searches with ATLAS and CMS at the LHC.  Figure~\ref{fig::sens} (left) displays the capability of a large LXe/LAr detector to reconstruct the WIMP mass and cross section for various masses and a hypothetical WIMP-nucleon cross section of 3$\times10^{-46}$ cm$^2$  \cite{Newstead:2013pea}. Exposures of 10\,t$\times$yr  and 20\,t$\times$yr using a LXe target only, as well as a combined exposure of 10\,t$\times$yr LXe and of 20\,t$\times$yr LAr, were assumed. The right side of the same Figure shows the reconstructed mass and cross sections for a 100\,GeV/$c^2$ WIMP, and several cross sections, using the combined exposure of 10\,t$\times$yr xenon  and 20\,t$\times$yr argon. The study assumes nuclear recoil energy thresholds of 6.6\,keV$_{\textnormal{\footnotesize nr}}$ and 20\,keV$_{\textnormal{\footnotesize nr}}$ in xenon and argon, respectively. The uncertainties on the dark matter halo parameters $\rho_0$=(0.3$\pm$0.1)\,GeV/cm$^2$, $v_0$=(220$\pm$20)\,km/s and $v_{esc}$=(544$\pm$40)\,km/s are taken into account.

\section{Summary and Outlook}

\begin{figure}[!h]
\centering
\includegraphics[scale=0.47]{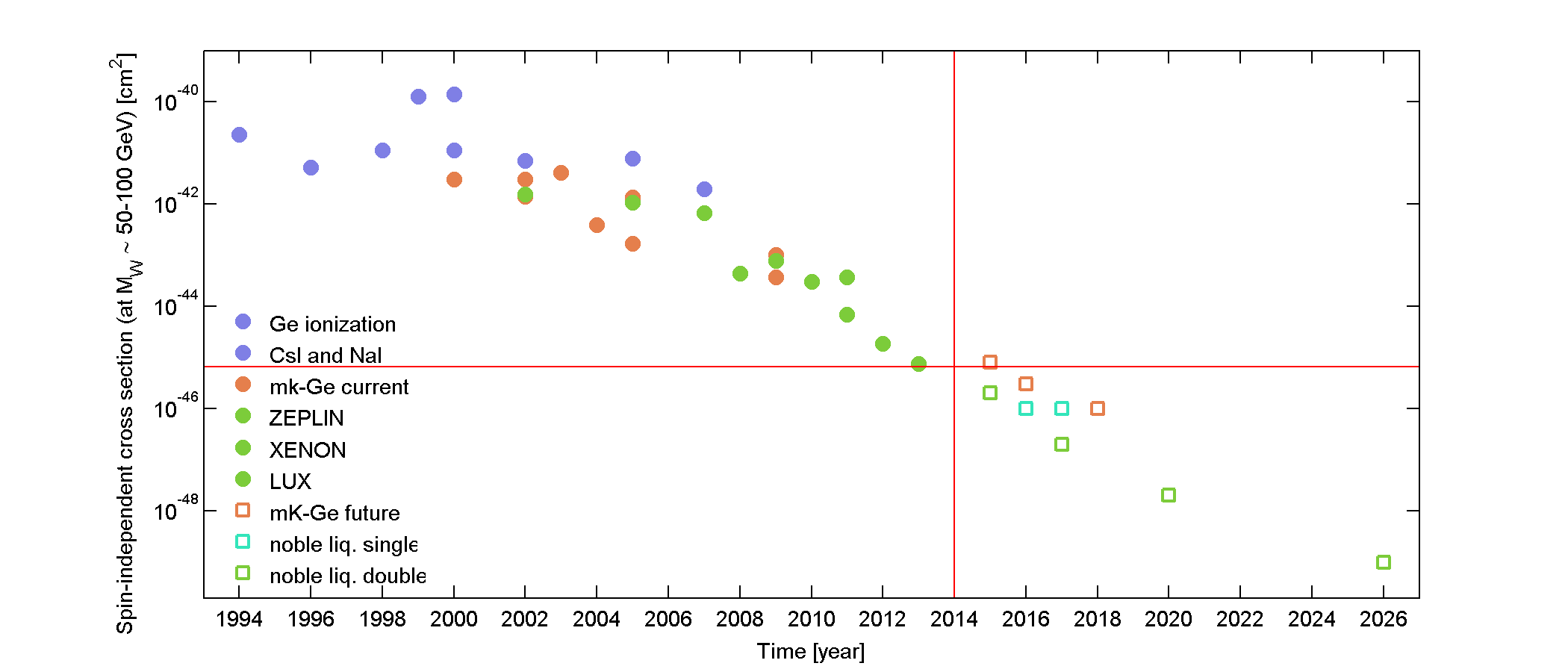}
\caption{\small{Existing upper limits on spin-independent WIMP-nucleon cross sections (for WIMP masses around 50-100\,GeV) from various direct detection techniques (filled circles), along with projections for the future (open squares), as a function time (Figure updated from \cite{Baudis:2012ig}).}}
\label{fig:si_limits_time}
\end{figure}

Dark matter detectors based on liquified noble gases have matured into a robust technology for detecting the minuscule energy deposited into a medium when a WIMP scatters off an atomic nucleus in the target. They have demonstrated the lowest backgrounds reached up to now in any direct detection experiment, and their scale up to large, homogeneous target masses is rather straightforward.  While detectors currently in operation have reached background levels as low as a few events/(keV\,t\,d) in the dark matter region of interest, before discrimination of electronic and nuclear recoils, detectors under commissioning and/or construction will improve these absolute backgrounds by a few orders of magnitude, until the irreducible neutrino background will start to dominate the recoil spectra. These ton- and multi-ton scale noble liquid detectors should reach sensitivities that will allow them to probe the entire experimentally accessible region for WIMP masses above $\sim$6\,GeV.  Figure\,\ref{fig:si_limits_time} shows the  evolution of upper limits on the spin-independent cross section as a function of time, together with latest projections for the future.

 \appendix
 
 \section*{Acknowledgments}
This work is supported by the University of Zurich, by the Swiss National Foundation (SNF) Grant No 200020-149256 and by the FP7 Marie Curie-ITN action PITN-GA-2011-289442 {\it Invisibles}.

\bibliographystyle{elsarticle-num}
\bibliography{baudis_taup2013}

\begin{thebibliography}{100}
\expandafter\ifx\csname url\endcsname\relax
  \def\url#1{\texttt{#1}}\fi
\expandafter\ifx\csname urlprefix\endcsname\relax\def\urlprefix{URL }\fi
\expandafter\ifx\csname href\endcsname\relax
  \def\href#1#2{#2} \def\path#1{#1}\fi

\bibitem{Ade:2013zuv}
P.~Ade, et~al., {Planck 2013 results. XVI. Cosmological parameters}\href
  {http://arxiv.org/abs/1303.5076} {\path{arXiv:1303.5076}}.

\bibitem{Lee:1977ua}
B.~W. Lee, S.~Weinberg, {Cosmological Lower Bound on Heavy Neutrino Masses},
  Phys.Rev.Lett. 39 (1977) 165--168.
\newblock \href {http://dx.doi.org/10.1103/PhysRevLett.39.165}
  {\path{doi:10.1103/PhysRevLett.39.165}}.

\bibitem{Jungman:1995df}
G.~Jungman, M.~Kamionkowski, K.~Griest, {Supersymmetric dark matter},
  Phys.Rept. 267 (1996) 195--373.
\newblock \href {http://arxiv.org/abs/hep-ph/9506380}
  {\path{arXiv:hep-ph/9506380}}, \href
  {http://dx.doi.org/10.1016/0370-1573(95)00058-5}
  {\path{doi:10.1016/0370-1573(95)00058-5}}.

\bibitem{Hooper:2007qk}
D.~Hooper, S.~Profumo, {Dark matter and collider phenomenology of universal
  extra dimensions}, Phys.Rept. 453 (2007) 29--115.
\newblock \href {http://arxiv.org/abs/hep-ph/0701197}
  {\path{arXiv:hep-ph/0701197}}, \href
  {http://dx.doi.org/10.1016/j.physrep.2007.09.003}
  {\path{doi:10.1016/j.physrep.2007.09.003}}.

\bibitem{Bertone:2004pz}
G.~Bertone, D.~Hooper, J.~Silk, {Particle dark matter: Evidence, candidates and
  constraints}, Phys.Rept. 405 (2005) 279--390.
\newblock \href {http://arxiv.org/abs/hep-ph/0404175}
  {\path{arXiv:hep-ph/0404175}}, \href
  {http://dx.doi.org/10.1016/j.physrep.2004.08.031}
  {\path{doi:10.1016/j.physrep.2004.08.031}}.

\bibitem{Raffelt:2002zz}
G.~Raffelt, {Axions}, Space Sci.Rev. 100 (2002) 153--158.
\newblock \href {http://dx.doi.org/10.1023/A:1015822212542}
  {\path{doi:10.1023/A:1015822212542}}.

\bibitem{Baker:2013zta}
K.~Baker, G.~Cantatore, S.~Cetin, M.~Davenport, K.~Desch, et~al., {The quest
  for axions and other new light particles}, Annalen Phys. 525 (2013) A93--A99.
\newblock \href {http://arxiv.org/abs/1306.2841} {\path{arXiv:1306.2841}},
  \href {http://dx.doi.org/10.1002/andp.201300727}
  {\path{doi:10.1002/andp.201300727}}.

\bibitem{Rybka:2014xca}
G.~Rybka, {Direct Detection Searches for Axion Dark Matter}, Phys.Dark Univ. 4
  (2014) 14--16.
\newblock \href {http://dx.doi.org/10.1016/j.dark.2014.05.003}
  {\path{doi:10.1016/j.dark.2014.05.003}}.

\bibitem{Goodman:1984dc}
M.~W. Goodman, E.~Witten, {Detectability of Certain Dark Matter Candidates},
  Phys.Rev. D31 (1985) 3059.
\newblock \href {http://dx.doi.org/10.1103/PhysRevD.31.3059}
  {\path{doi:10.1103/PhysRevD.31.3059}}.

\bibitem{Lewin:1995rx}
J.~Lewin, P.~Smith, {Review of mathematics, numerical factors, and corrections
  for dark matter experiments based on elastic nuclear recoil}, Astropart.Phys.
  6 (1996) 87--112.
\newblock \href {http://dx.doi.org/10.1016/S0927-6505(96)00047-3}
  {\path{doi:10.1016/S0927-6505(96)00047-3}}.

\bibitem{Green:2011bv}
A.~M. Green, {Astrophysical uncertainties on direct detection experiments},
  Mod.Phys.Lett. A27 (2012) 1230004.
\newblock \href {http://arxiv.org/abs/1112.0524} {\path{arXiv:1112.0524}},
  \href {http://dx.doi.org/10.1142/S0217732312300042}
  {\path{doi:10.1142/S0217732312300042}}.

\bibitem{Smith:2006ym}
M.~C. Smith, et~al., {The RAVE Survey: Constraining the Local Galactic Escape
  Speed}, Mon. Not. Roy. Astron. Soc. 379 (2007) 755--772.
\newblock \href {http://arxiv.org/abs/astro-ph/0611671}
  {\path{arXiv:astro-ph/0611671}}, \href
  {http://dx.doi.org/10.1111/j.1365-2966.2007.11964.x}
  {\path{doi:10.1111/j.1365-2966.2007.11964.x}}.

\bibitem{Vogelsberger:2008qb}
M.~Vogelsberger, A.~Helmi, V.~Springel, S.~D. White, J.~Wang, et~al.,
  {Phase-space structure in the local dark matter distribution and its
  signature in direct detection experiments}, Mon.Not.Roy.Astron.Soc. 395
  (2009) 797--811.
\newblock \href {http://arxiv.org/abs/0812.0362} {\path{arXiv:0812.0362}},
  \href {http://dx.doi.org/10.1111/j.1365-2966.2009.14630.x}
  {\path{doi:10.1111/j.1365-2966.2009.14630.x}}.

\bibitem{Read:2014qva}
J.~Read, {The Local Dark Matter Density}\href {http://arxiv.org/abs/1404.1938}
  {\path{arXiv:1404.1938}}.

\bibitem{Menendez:2012tm}
J.~Menendez, D.~Gazit, A.~Schwenk, {Spin-dependent WIMP scattering off nuclei},
  Phys.Rev. D86 (2012) 103511.
\newblock \href {http://arxiv.org/abs/1208.1094} {\path{arXiv:1208.1094}},
  \href {http://dx.doi.org/10.1103/PhysRevD.86.103511}
  {\path{doi:10.1103/PhysRevD.86.103511}}.

\bibitem{Klos:2013rwa}
P.~Klos, J.~MenŽndez, D.~Gazit, A.~Schwenk, {Large-scale nuclear structure
  calculations for spin-dependent WIMP scattering with chiral effective field
  theory currents}, Physical Review D 88, 083516.
\newblock \href {http://arxiv.org/abs/1304.7684} {\path{arXiv:1304.7684}}.

\bibitem{Fan:2010gt}
J.~Fan, M.~Reece, L.-T. Wang, {Non-relativistic effective theory of dark matter
  direct detection}, JCAP 1011 (2010) 042.
\newblock \href {http://arxiv.org/abs/1008.1591} {\path{arXiv:1008.1591}},
  \href {http://dx.doi.org/10.1088/1475-7516/2010/11/042}
  {\path{doi:10.1088/1475-7516/2010/11/042}}.

\bibitem{Fitzpatrick:2012ix}
A.~L. Fitzpatrick, W.~Haxton, E.~Katz, N.~Lubbers, Y.~Xu, {The Effective Field
  Theory of Dark Matter Direct Detection}, JCAP 1302 (2013) 004.
\newblock \href {http://arxiv.org/abs/1203.3542} {\path{arXiv:1203.3542}},
  \href {http://dx.doi.org/10.1088/1475-7516/2013/02/004}
  {\path{doi:10.1088/1475-7516/2013/02/004}}.

\bibitem{Hill:2013hoa}
R.~J. Hill, M.~P. Solon, {WIMP-nucleon scattering with heavy WIMP effective
  theory}, Phys.Rev.Lett. 112 (2014) 211602.
\newblock \href {http://arxiv.org/abs/1309.4092} {\path{arXiv:1309.4092}},
  \href {http://dx.doi.org/10.1103/PhysRevLett.112.211602}
  {\path{doi:10.1103/PhysRevLett.112.211602}}.

\bibitem{Anand:2013yka}
N.~Anand, A.~L. Fitzpatrick, W.~Haxton, {Model-independent WIMP Scattering
  Responses and Event Rates: A Mathematica Package for Experimental Analysis},
  Phys.Rev. C89 (2014) 065501.
\newblock \href {http://arxiv.org/abs/1308.6288} {\path{arXiv:1308.6288}},
  \href {http://dx.doi.org/10.1103/PhysRevC.89.065501}
  {\path{doi:10.1103/PhysRevC.89.065501}}.

\bibitem{Baudis:2013bba}
L.~Baudis, G.~Kessler, P.~Klos, R.~Lang, J.~Menendez, et~al., {Signatures of
  Dark Matter Scattering Inelastically Off Nuclei}, Phys.Rev. D88 (2013)
  115014.
\newblock \href {http://arxiv.org/abs/1309.0825} {\path{arXiv:1309.0825}},
  \href {http://dx.doi.org/10.1103/PhysRevD.88.115014}
  {\path{doi:10.1103/PhysRevD.88.115014}}.

\bibitem{Heusser:1995wd}
G.~Heusser, {Low-radioactivity background techniques}, Ann.Rev.Nucl.Part.Sci.
  45 (1995) 543--590.
\newblock \href {http://dx.doi.org/10.1146/annurev.ns.45.120195.002551}
  {\path{doi:10.1146/annurev.ns.45.120195.002551}}.

\bibitem{Haffke:2011fp}
M.~Haffke, L.~Baudis, T.~Bruch, A.~Ferella, T.~Marrodan~Undagoitia, et~al.,
  {Background Measurements in the Gran Sasso Underground Laboratory},
  Nucl.Instrum.Meth. A643 (2011) 36--41.
\newblock \href {http://arxiv.org/abs/1101.5298} {\path{arXiv:1101.5298}},
  \href {http://dx.doi.org/10.1016/j.nima.2011.04.027}
  {\path{doi:10.1016/j.nima.2011.04.027}}.

\bibitem{Selvi:2011zz}
M.~Selvi, {Study of the performances of the shield and muon veto of the XENON1T
  experiment}, PoS IDM2010 (2011) 053.

\bibitem{Baudis:2011am}
L.~Baudis, A.~Ferella, A.~Askin, J.~Angle, E.~Aprile, et~al., {Gator: a
  low-background counting facility at the Gran Sasso Underground Laboratory},
  JINST 6 (2011) P08010.
\newblock \href {http://arxiv.org/abs/1103.2125} {\path{arXiv:1103.2125}},
  \href {http://dx.doi.org/10.1088/1748-0221/6/08/P08010}
  {\path{doi:10.1088/1748-0221/6/08/P08010}}.

\bibitem{Mei:2008ir}
D.-M. Mei, C.~Zhang, A.~Hime, {Evaluation of (alpha,n) Induced Neutrons as a
  Background for Dark Matter Experiments}, Nucl.Instrum.Meth. A606 (2009)
  651--660.
\newblock \href {http://arxiv.org/abs/0812.4307} {\path{arXiv:0812.4307}},
  \href {http://dx.doi.org/10.1016/j.nima.2009.04.032}
  {\path{doi:10.1016/j.nima.2009.04.032}}.

\bibitem{Baudis:2013qla}
L.~Baudis, A.~Ferella, A.~Kish, A.~Manalaysay, T.~M. Undagoitia, et~al.,
  {Neutrino physics with multi-ton scale liquid xenon detectors}, JCAP 01
  (2014) 044.
\newblock \href {http://arxiv.org/abs/1309.7024} {\path{arXiv:1309.7024}},
  \href {http://dx.doi.org/10.1088/1475-7516/2014/01/044}
  {\path{doi:10.1088/1475-7516/2014/01/044}}.

\bibitem{Strigari:2009bq}
L.~E. Strigari, {Neutrino Coherent Scattering Rates at Direct Dark Matter
  Detectors}, New J.Phys. 11 (2009) 105011.
\newblock \href {http://arxiv.org/abs/0903.3630} {\path{arXiv:0903.3630}},
  \href {http://dx.doi.org/10.1088/1367-2630/11/10/105011}
  {\path{doi:10.1088/1367-2630/11/10/105011}}.

\bibitem{Gutlein:2010tq}
A.~Gutlein, C.~Ciemniak, F.~von Feilitzsch, N.~Haag, M.~Hofmann, et~al., {Solar
  and atmospheric neutrinos: Background sources for the direct dark matter
  search}, Astropart.Phys. 34 (2010) 90--96.
\newblock \href {http://arxiv.org/abs/1003.5530} {\path{arXiv:1003.5530}},
  \href {http://dx.doi.org/10.1016/j.astropartphys.2010.06.002}
  {\path{doi:10.1016/j.astropartphys.2010.06.002}}.

\bibitem{Billard:2013qya}
J.~Billard, L.~Strigari, E.~Figueroa-Feliciano, {Implication of neutrino
  backgrounds on the reach of next generation dark matter direct detection
  experiments}, Phys.Rev. D89 (2014) 023524.
\newblock \href {http://arxiv.org/abs/1307.5458} {\path{arXiv:1307.5458}}.

\bibitem{Anderson:2011bi}
A.~Anderson, J.~Conrad, E.~Figueroa-Feliciano, K.~Scholberg, J.~Spitz,
  {Coherent Neutrino Scattering in Dark Matter Detectors}, Phys.Rev. D84 (2011)
  013008.
\newblock \href {http://arxiv.org/abs/1103.4894} {\path{arXiv:1103.4894}},
  \href {http://dx.doi.org/10.1103/PhysRevD.84.013008}
  {\path{doi:10.1103/PhysRevD.84.013008}}.

\bibitem{Lippincott:2011zr}
W.~Lippincott, et~al., {Scintillation yield and time dependence from electronic
  and nuclear recoils in liquid neon}, Phys.Rev. C86 (2012) 015807.
\newblock \href {http://arxiv.org/abs/1111.3260} {\path{arXiv:1111.3260}},
  \href {http://dx.doi.org/10.1103/PhysRevC.86.015807}
  {\path{doi:10.1103/PhysRevC.86.015807}}.

\bibitem{Aprile:1900zz}
E.~Aprile, L.~Baudis, {Liquid noble gases, In {\sl Bertone, G. (ed.): Particle
  dark matter} 413-436}.

\bibitem{Chepel:2012sj}
V.~Chepel, H.~Araujo, {Liquid noble gas detectors for low energy particle
  physics}, JINST 8 (2013) R04001.
\newblock \href {http://arxiv.org/abs/1207.2292} {\path{arXiv:1207.2292}},
  \href {http://dx.doi.org/10.1088/1748-0221/8/04/R04001}
  {\path{doi:10.1088/1748-0221/8/04/R04001}}.

\bibitem{Albert:2013gpz}
J.~Albert, et~al., {An improved measurement of the $2\nu \beta \beta\ $
  half-life of Xe-136 with EXO-200}, Phys.Rev. C89 (2014) 015502.
\newblock \href {http://arxiv.org/abs/1306.6106} {\path{arXiv:1306.6106}},
  \href {http://dx.doi.org/10.1103/PhysRevC.89.015502}
  {\path{doi:10.1103/PhysRevC.89.015502}}.

\bibitem{Benetti:2006az}
P.~Benetti, et~al., {Measurement of the specific activity of ar-39 in natural
  argon}, Nucl.Instrum.Meth. A574 (2007) 83--88.
\newblock \href {http://arxiv.org/abs/astro-ph/0603131}
  {\path{arXiv:astro-ph/0603131}}, \href
  {http://dx.doi.org/10.1016/j.nima.2007.01.106}
  {\path{doi:10.1016/j.nima.2007.01.106}}.

\bibitem{Back:2012pg}
H.~O. Back, F.~Calaprice, C.~Condon, E.~de~Haas, R.~Ford, et~al., {First Large
  Scale Production of Low Radioactivity Argon From Underground Sources}\href
  {http://arxiv.org/abs/1204.6024} {\path{arXiv:1204.6024}}.

\bibitem{Lindemann:2013kna}
S.~Lindemann, H.~Simgen, {Krypton assay in xenon at the ppq level using a gas
  chromatographic system and mass spectrometer}, Eur.Phys.J. C74 (2014) 2746.
\newblock \href {http://arxiv.org/abs/1308.4806} {\path{arXiv:1308.4806}},
  \href {http://dx.doi.org/10.1140/epjc/s10052-014-2746-1}
  {\path{doi:10.1140/epjc/s10052-014-2746-1}}.

\bibitem{Akerib:2013tjd}
D.~Akerib, et~al., {First results from the LUX dark matter experiment at the
  Sanford Underground Research Facility}\href {http://arxiv.org/abs/1310.8214}
  {\path{arXiv:1310.8214}}.

\bibitem{Abe:2013tc}
K.~Abe, K.~Hieda, K.~Hiraide, S.~Hirano, Y.~Kishimoto, et~al., {XMASS
  detector}, Nucl.Instrum.Meth. A716 (2013) 78--85.
\newblock \href {http://arxiv.org/abs/1301.2815} {\path{arXiv:1301.2815}},
  \href {http://dx.doi.org/10.1016/j.nima.2013.03.059}
  {\path{doi:10.1016/j.nima.2013.03.059}}.

\bibitem{Hitachi:1983zz}
A.~Hitachi, T.~Takahashi, N.~Funayama, K.~Masuda, J.~Kikuchi, et~al., {Effect
  of ionization density on the time dependence of luminescence from liquid
  argon and xenon}, Phys.Rev. B27 (1983) 5279--5285.
\newblock \href {http://dx.doi.org/10.1103/PhysRevB.27.5279}
  {\path{doi:10.1103/PhysRevB.27.5279}}.

\bibitem{Lippincott:2008ad}
W.~Lippincott, K.~Coakley, D.~Gastler, A.~Hime, E.~Kearns, et~al.,
  {Scintillation time dependence and pulse shape discrimination in liquid
  argon}, Phys.Rev. C78 (2008) 035801.
\newblock \href {http://arxiv.org/abs/0801.1531} {\path{arXiv:0801.1531}},
  \href {http://dx.doi.org/10.1103/PhysRevC.81.039901,
  10.1103/PhysRevC.78.035801} {\path{doi:10.1103/PhysRevC.81.039901,
  10.1103/PhysRevC.78.035801}}.

\bibitem{Sorensen:2011bd}
P.~Sorensen, C.~E. Dahl, {Nuclear recoil energy scale in liquid xenon with
  application to the direct detection of dark matter}, Phys.Rev. D83 (2011)
  063501.
\newblock \href {http://arxiv.org/abs/1101.6080} {\path{arXiv:1101.6080}},
  \href {http://dx.doi.org/10.1103/PhysRevD.83.063501}
  {\path{doi:10.1103/PhysRevD.83.063501}}.

\bibitem{Aprile:2013blg}
E.~Aprile, et~al., {Observation and applications of single-electron charge
  signals in the XENON100 experiment}, J.Phys. G41 (2014) 035201.
\newblock \href {http://arxiv.org/abs/1311.1088} {\path{arXiv:1311.1088}},
  \href {http://dx.doi.org/10.1088/0954-3899/41/3/035201}
  {\path{doi:10.1088/0954-3899/41/3/035201}}.

\bibitem{Szydagis:2014xog}
M.~Szydagis, et~al., {A Detailed Look at the First Results from the Large
  Underground Xenon (LUX) Dark Matter Experiment}\href
  {http://arxiv.org/abs/1402.3731} {\path{arXiv:1402.3731}}.

\bibitem{Manalaysay:2009yq}
A.~Manalaysay, T.~Undagoitia, A.~Askin, L.~Baudis, A.~Behrens, et~al.,
  {Spatially uniform calibration of a liquid xenon detector at low energies
  using 83m-Kr}, Rev.Sci.Instrum. 81 (2010) 073303.
\newblock \href {http://arxiv.org/abs/0908.0616} {\path{arXiv:0908.0616}},
  \href {http://dx.doi.org/10.1063/1.3436636} {\path{doi:10.1063/1.3436636}}.

\bibitem{Kastens:2009pa}
L.~Kastens, S.~Cahn, A.~Manzur, D.~McKinsey, {Calibration of a Liquid Xenon
  Detector with Kr-83m}, Phys.Rev. C80 (2009) 045809.
\newblock \href {http://arxiv.org/abs/0905.1766} {\path{arXiv:0905.1766}},
  \href {http://dx.doi.org/10.1103/PhysRevC.80.045809}
  {\path{doi:10.1103/PhysRevC.80.045809}}.

\bibitem{Baudis:2013cca}
L.~Baudis, H.~Dujmovic, C.~Geis, A.~James, A.~Kish, et~al., {Response of liquid
  xenon to Compton electrons down to 1.5 keV}, Phys.Rev. D87~(11) (2013)
  115015.
\newblock \href {http://arxiv.org/abs/1303.6891} {\path{arXiv:1303.6891}},
  \href {http://dx.doi.org/10.1103/PhysRevD.87.115015}
  {\path{doi:10.1103/PhysRevD.87.115015}}.

\bibitem{Szydagis:2013sih}
M.~Szydagis, A.~Fyhrie, D.~Thorngren, M.~Tripathi, {Enhancement of NEST
  Capabilities for Simulating Low-Energy Recoils in Liquid Xenon}, JINST 8
  (2013) C10003.
\newblock \href {http://arxiv.org/abs/1307.6601} {\path{arXiv:1307.6601}},
  \href {http://dx.doi.org/10.1088/1748-0221/8/10/C10003}
  {\path{doi:10.1088/1748-0221/8/10/C10003}}.

\bibitem{Szydagis:2011tk}
M.~Szydagis, N.~Barry, K.~Kazkaz, J.~Mock, D.~Stolp, et~al., {NEST: A
  Comprehensive Model for Scintillation Yield in Liquid Xenon}, JINST 6 (2011)
  P10002.
\newblock \href {http://arxiv.org/abs/1106.1613} {\path{arXiv:1106.1613}},
  \href {http://dx.doi.org/10.1088/1748-0221/6/10/P10002}
  {\path{doi:10.1088/1748-0221/6/10/P10002}}.

\bibitem{Plante:2011hw}
G.~Plante, E.~Aprile, R.~Budnik, B.~Choi, K.~Giboni, et~al., {New Measurement
  of the Scintillation Efficiency of Low-Energy Nuclear Recoils in Liquid
  Xenon}, Phys.Rev. C84 (2011) 045805.
\newblock \href {http://arxiv.org/abs/1104.2587} {\path{arXiv:1104.2587}},
  \href {http://dx.doi.org/10.1103/PhysRevC.84.045805}
  {\path{doi:10.1103/PhysRevC.84.045805}}.

\bibitem{Aprile:2013teh}
E.~Aprile, et~al., {Response of the XENON100 Dark Matter Detector to Nuclear
  Recoils}, Phys.Rev. D88 (2013) 012006.
\newblock \href {http://arxiv.org/abs/1304.1427} {\path{arXiv:1304.1427}},
  \href {http://dx.doi.org/10.1103/PhysRevD.88.012006}
  {\path{doi:10.1103/PhysRevD.88.012006}}.

\bibitem{lux_ucla2014}
C.~Carmona, {First dark matter results from the LUX experiment}, Talk at DM
  2014, UCLA.

\bibitem{Gastler:2010sc}
D.~Gastler, E.~Kearns, A.~Hime, L.~Stonehill, S.~Seibert, et~al., {Measurement
  of scintillation efficiency for nuclear recoils in liquid argon}, Phys.Rev.
  C85 (2012) 065811.
\newblock \href {http://arxiv.org/abs/1004.0373} {\path{arXiv:1004.0373}},
  \href {http://dx.doi.org/10.1103/PhysRevC.85.065811}
  {\path{doi:10.1103/PhysRevC.85.065811}}.

\bibitem{Regenfus:2012kh}
C.~Regenfus, Y.~Allkofer, C.~Amsler, W.~Creus, A.~Ferella, et~al., {Study of
  nuclear recoils in liquid argon with monoenergetic neutrons},
  J.Phys.Conf.Ser. 375 (2012) 012019.
\newblock \href {http://arxiv.org/abs/1203.0849} {\path{arXiv:1203.0849}},
  \href {http://dx.doi.org/10.1088/1742-6596/375/1/012019}
  {\path{doi:10.1088/1742-6596/375/1/012019}}.

\bibitem{Alexander:2013aaq}
T.~Alexander, et~al., {Observation of the Dependence of Scintillation from
  Nuclear Recoils in Liquid Argon on Drift Field}, Phys.Rev. D88 (2013) 092006.
\newblock \href {http://arxiv.org/abs/1306.5675} {\path{arXiv:1306.5675}},
  \href {http://dx.doi.org/10.1103/PhysRevD.88.092006}
  {\path{doi:10.1103/PhysRevD.88.092006}}.

\bibitem{Lippincot_lhcforum_2014}
H.~Lippincot, {Response of liquid noble gases to nuclear recoils}, Talk at the
  LHC results forum.

\bibitem{Aprile:2006kx}
E.~Aprile, et~al., {Simultaneous Measurement of Ionization and Scintillation
  from Nuclear Recoils in Liquid Xenon as Target for a Dark Matter Experiment},
  Phys. Rev. Lett. 97 (2006) 081302.
\newblock \href {http://arxiv.org/abs/astro-ph/0601552}
  {\path{arXiv:astro-ph/0601552}}, \href
  {http://dx.doi.org/10.1103/PhysRevLett.97.081302}
  {\path{doi:10.1103/PhysRevLett.97.081302}}.

\bibitem{Manzur:2009hp}
A.~Manzur, A.~Curioni, L.~Kastens, D.~McKinsey, K.~Ni, et~al., {Scintillation
  efficiency and ionization yield of liquid xenon for mono-energetic nuclear
  recoils down to 4 keV}, Phys.Rev. C81 (2010) 025808.
\newblock \href {http://arxiv.org/abs/0909.1063} {\path{arXiv:0909.1063}},
  \href {http://dx.doi.org/10.1103/PhysRevC.81.025808}
  {\path{doi:10.1103/PhysRevC.81.025808}}.

\bibitem{Bezrukov:2010qa}
F.~Bezrukov, F.~Kahlhoefer, M.~Lindner, {Interplay between scintillation and
  ionization in liquid xenon Dark Matter searches}, Astropart.Phys. 35 (2011)
  119--127.
\newblock \href {http://arxiv.org/abs/1011.3990} {\path{arXiv:1011.3990}},
  \href {http://dx.doi.org/10.1016/j.astropartphys.2011.06.008}
  {\path{doi:10.1016/j.astropartphys.2011.06.008}}.

\bibitem{Grandi:2005dm}
L.~Grandi, {WARP: an argon double phase technique for Dark Matter search},
  Ph.D. Thesis, University of Pavia.

\bibitem{Benetti:2007cd}
P.~Benetti, et~al., {First results from a dark matter search with liquid argon
  at 87-K in the Gran Sasso underground laboratory}, Astropart. Phys. 28 (2008)
  495--507.
\newblock \href {http://arxiv.org/abs/astro-ph/0701286}
  {\path{arXiv:astro-ph/0701286}}, \href
  {http://dx.doi.org/10.1016/j.astropartphys.2007.08.002}
  {\path{doi:10.1016/j.astropartphys.2007.08.002}}.

\bibitem{Cao:2014gns}
H.~Cao, et~al., {Measurement of Scintillation and Ionization Yield and
  Scintillation Pulse Shape from Nuclear Recoils in Liquid Argon}\href
  {http://arxiv.org/abs/1406.4825} {\path{arXiv:1406.4825}}.

\bibitem{Aprile:2012an}
E.~Aprile, R.~Budnik, B.~Choi, H.~Contreras, K.~Giboni, et~al., {Measurement of
  the Scintillation Yield of Low-Energy Electrons in Liquid Xenon}, Phys.Rev.
  D86 (2012) 112004.
\newblock \href {http://arxiv.org/abs/1209.3658} {\path{arXiv:1209.3658}},
  \href {http://dx.doi.org/10.1103/PhysRevD.86.112004}
  {\path{doi:10.1103/PhysRevD.86.112004}}.

\bibitem{Abe:2012ut}
K.~Abe, K.~Hieda, K.~Hiraide, S.~Hirano, Y.~Kishimoto, et~al., {Search for
  solar axions in XMASS, a large liquid-xenon detector}, Phys. Lett. B 724
  (2013) 46--50.
\newblock \href {http://arxiv.org/abs/1212.6153} {\path{arXiv:1212.6153}},
  \href {http://dx.doi.org/10.1016/j.physletb.2013.05.060}
  {\path{doi:10.1016/j.physletb.2013.05.060}}.

\bibitem{Aprile:2014eoa}
E.~Aprile, F.~Agostini, M.~Alfonsi, K.~Arisaka, F.~Arneodo, et~al., {First
  Axion Results from the XENON100 Experiment}\href
  {http://arxiv.org/abs/1404.1455} {\path{arXiv:1404.1455}}.

\bibitem{Peiffer:2008zz}
P.~Peiffer, T.~Pollmann, S.~Schonert, A.~Smolnikov, S.~Vasiliev, {Pulse shape
  analysis of scintillation signals from pure and xenon-doped liquid argon for
  radioactive background identification}, JINST 3 (2008) P08007.
\newblock \href {http://dx.doi.org/10.1088/1748-0221/3/08/P08007}
  {\path{doi:10.1088/1748-0221/3/08/P08007}}.

\bibitem{Amsler:2007gs}
C.~Amsler, et~al., {Luminescence quenching of the triplet excimer state by air
  traces in gaseous argon}, JINST 3 (2008) P02001.
\newblock \href {http://arxiv.org/abs/0708.2621} {\path{arXiv:0708.2621}},
  \href {http://dx.doi.org/10.1088/1748-0221/3/02/P02001}
  {\path{doi:10.1088/1748-0221/3/02/P02001}}.

\bibitem{Akimov:2011tj}
D.~Y. Akimov, H.~Araujo, E.~Barnes, V.~Belov, A.~Bewick, et~al., {WIMP-nucleon
  cross-section results from the second science run of ZEPLIN-III}, Phys.Lett.
  B709 (2012) 14--20.
\newblock \href {http://arxiv.org/abs/1110.4769} {\path{arXiv:1110.4769}},
  \href {http://dx.doi.org/10.1016/j.physletb.2012.01.064}
  {\path{doi:10.1016/j.physletb.2012.01.064}}.

\bibitem{Akimov:2010zz}
D.~Y. Akimov, {The ZEPLIN-III dark matter detector}, Nucl.Instrum.Meth. A623
  (2010) 451--453.
\newblock \href {http://dx.doi.org/10.1016/j.nima.2010.03.033}
  {\path{doi:10.1016/j.nima.2010.03.033}}.

\bibitem{Angle:2007uj}
J.~Angle, et~al., {First Results from the XENON10 Dark Matter Experiment at the
  Gran Sasso National Laboratory}, Phys.Rev.Lett. 100 (2008) 021303.
\newblock \href {http://arxiv.org/abs/0706.0039} {\path{arXiv:0706.0039}},
  \href {http://dx.doi.org/10.1103/PhysRevLett.100.021303}
  {\path{doi:10.1103/PhysRevLett.100.021303}}.

\bibitem{Aprile:2010bt}
E.~Aprile, et~al., {Design and Performance of the XENON10 Dark Matter
  Experiment}, Astropart.Phys. 34 (2011) 679--698.
\newblock \href {http://arxiv.org/abs/1001.2834} {\path{arXiv:1001.2834}},
  \href {http://dx.doi.org/10.1016/j.astropartphys.2011.01.006}
  {\path{doi:10.1016/j.astropartphys.2011.01.006}}.

\bibitem{Aprile:2011hi}
E.~Aprile, et~al., {Dark Matter Results from 100 Live Days of XENON100 Data},
  Phys.Rev.Lett. 107 (2011) 131302.
\newblock \href {http://arxiv.org/abs/1104.2549} {\path{arXiv:1104.2549}},
  \href {http://dx.doi.org/10.1103/PhysRevLett.107.131302}
  {\path{doi:10.1103/PhysRevLett.107.131302}}.

\bibitem{Aprile:2011dd}
E.~Aprile, et~al., {The XENON100 Dark Matter Experiment}, Astropart.Phys. 35
  (2012) 573--590.
\newblock \href {http://arxiv.org/abs/1107.2155} {\path{arXiv:1107.2155}},
  \href {http://dx.doi.org/10.1016/j.astropartphys.2012.01.003}
  {\path{doi:10.1016/j.astropartphys.2012.01.003}}.

\bibitem{Aprile:2012nq}
E.~Aprile, et~al., {Dark Matter Results from 225 Live Days of XENON100
  Data}\href {http://arxiv.org/abs/1207.5988} {\path{arXiv:1207.5988}}.

\bibitem{Aprile:2012vw}
E.~Aprile, et~al., {Analysis of the XENON100 Dark Matter Search Data},
  Astropart.Phys. 54 (2014) 11--24.
\newblock \href {http://arxiv.org/abs/1207.3458} {\path{arXiv:1207.3458}},
  \href {http://dx.doi.org/10.1016/j.astropartphys.2013.10.002}
  {\path{doi:10.1016/j.astropartphys.2013.10.002}}.

\bibitem{Aprile:2013doa}
E.~Aprile, et~al., {Limits on spin-dependent WIMP-nucleon cross sections from
  225 live days of XENON100 data}, Phys.Rev.Lett. 111~(2) (2013) 021301.
\newblock \href {http://arxiv.org/abs/1301.6620} {\path{arXiv:1301.6620}},
  \href {http://dx.doi.org/10.1103/PhysRevLett.111.021301}
  {\path{doi:10.1103/PhysRevLett.111.021301}}.

\bibitem{Akerib:2012ys}
D.~Akerib, et~al., {The Large Underground Xenon (LUX) Experiment},
  Nucl.Instrum.Meth. A704 (2013) 111--126.
\newblock \href {http://arxiv.org/abs/1211.3788} {\path{arXiv:1211.3788}},
  \href {http://dx.doi.org/10.1016/j.nima.2012.11.135}
  {\path{doi:10.1016/j.nima.2012.11.135}}.

\bibitem{Moriyama:2011zz}
S.~Moriyama, {Status of XMASS experiment}, PoS IDM2010 (2011) 057.

\bibitem{Abe:2012az}
K.~Abe, et~al., {Light WIMP search in XMASS}, Phys.Lett. B719 (2013) 78--82.
\newblock \href {http://arxiv.org/abs/1211.5404} {\path{arXiv:1211.5404}},
  \href {http://dx.doi.org/10.1016/j.physletb.2013.01.001}
  {\path{doi:10.1016/j.physletb.2013.01.001}}.

\bibitem{Uchida:2014cnn}
H.~Uchida, et~al., {Search for inelastic WIMP nucleus scattering on $^{129}$Xe
  in data from the XMASS-I experiment}\href {http://arxiv.org/abs/1401.4737}
  {\path{arXiv:1401.4737}}.

\bibitem{Cao:2014jsa}
X.~Cao, X.~Chen, Y.~Chen, X.~Cui, D.~Fang, et~al., {PandaX: A Liquid Xenon Dark
  Matter Experiment at CJPL}\href {http://arxiv.org/abs/1405.2882}
  {\path{arXiv:1405.2882}}.

\bibitem{pandax_ucla2014}
X.~Ji, {Status of PandaX}, Talk at DM 2014, UCLA.

\bibitem{Baudis:2013xva}
L.~Baudis, A.~Behrens, A.~Ferella, A.~Kish, T.~Marrodan~Undagoitia, et~al.,
  {Performance of the Hamamatsu R11410 Photomultiplier Tube in cryogenic Xenon
  Environments}, JINST 8 (2013) P04026.
\newblock \href {http://arxiv.org/abs/1303.0226} {\path{arXiv:1303.0226}},
  \href {http://dx.doi.org/10.1088/1748-0221/8/04/P04026}
  {\path{doi:10.1088/1748-0221/8/04/P04026}}.

\bibitem{Bossa:2014cfa}
M.~Bossa, {DarkSide-50, a background free experiment for dark matter searches},
  JINST 9 (2014) C01034.
\newblock \href {http://dx.doi.org/10.1088/1748-0221/9/01/C01034}
  {\path{doi:10.1088/1748-0221/9/01/C01034}}.

\bibitem{grandi_ucla2014}
L.~Grandi, {DarkSide-50}, Talk at DM 2014, UCLA.

\bibitem{Marchionni:2010fi}
A.~Marchionni, et~al., {ArDM: a ton-scale LAr detector for direct Dark Matter
  searches}, J.Phys.Conf.Ser. 308 (2011) 012006.
\newblock \href {http://arxiv.org/abs/1012.5967} {\path{arXiv:1012.5967}},
  \href {http://dx.doi.org/10.1088/1742-6596/308/1/012006}
  {\path{doi:10.1088/1742-6596/308/1/012006}}.

\bibitem{Badertscher:2013ygt}
A.~Badertscher, F.~Bay, N.~Bourgeois, C.~Cantini, A.~Curioni, et~al., {ArDM:
  first results from underground commissioning}, JINST 8 (2013) C09005.
\newblock \href {http://arxiv.org/abs/1309.3992} {\path{arXiv:1309.3992}},
  \href {http://dx.doi.org/10.1088/1748-0221/8/09/C09005}
  {\path{doi:10.1088/1748-0221/8/09/C09005}}.

\bibitem{Rielage:2014pfm}
K.~Rielage, et~al., {Update on the MiniCLEAN Dark Matter Experiment}\href
  {http://arxiv.org/abs/1403.4842} {\path{arXiv:1403.4842}}.

\bibitem{Boulay:2012hq}
M.~Boulay, {DEAP-3600 Dark Matter Search at SNOLAB}, J.Phys.Conf.Ser. 375
  (2012) 012027.
\newblock \href {http://arxiv.org/abs/1203.0604} {\path{arXiv:1203.0604}},
  \href {http://dx.doi.org/10.1088/1742-6596/375/1/012027}
  {\path{doi:10.1088/1742-6596/375/1/012027}}.

\bibitem{Cushman:2013zza}
P.~Cushman, C.~Galbiati, D.~McKinsey, H.~Robertson, T.~Tait, et~al., {Snowmass
  CF1 Summary: WIMP Dark Matter Direct Detection}\href
  {http://arxiv.org/abs/1310.8327} {\path{arXiv:1310.8327}}.

\bibitem{Malling:2011va}
D.~Malling, D.~Akerib, H.~Araujo, X.~Bai, S.~Bedikian, et~al., {After LUX: The
  LZ Program}\href {http://arxiv.org/abs/1110.0103} {\path{arXiv:1110.0103}}.

\bibitem{Baudis:2012bc}
L.~Baudis, {DARWIN: dark matter WIMP search with noble liquids},
  J.Phys.Conf.Ser. 375 (2012) 012028.
\newblock \href {http://arxiv.org/abs/1201.2402} {\path{arXiv:1201.2402}},
  \href {http://dx.doi.org/10.1088/1742-6596/375/1/012028}
  {\path{doi:10.1088/1742-6596/375/1/012028}}.

\bibitem{Agnese:2014aze}
R.~Agnese, et~al., {Search for Low-Mass WIMPs with SuperCDMS}\href
  {http://arxiv.org/abs/1402.7137} {\path{arXiv:1402.7137}}.

\bibitem{Kraus:2011zz}
H.~Kraus, E.~Armengaud, C.~Augier, M.~Bauer, N.~Bechtold, et~al., {EURECA}, PoS
  IDM2010 (2011) 109.

\bibitem{Alexander:2013hia}
T.~Alexander, et~al., {DarkSide search for dark matter}, JINST 8 (2013) C11021.
\newblock \href {http://dx.doi.org/10.1088/1748-0221/8/11/C11021}
  {\path{doi:10.1088/1748-0221/8/11/C11021}}.

\bibitem{Baudis:2010ch}
L.~Baudis, {DARWIN: dark matter WIMP search with noble liquids}, PoS IDM2010
  (2011) 122.
\newblock \href {http://arxiv.org/abs/1012.4764} {\path{arXiv:1012.4764}}.

\bibitem{Pato:2010zk}
M.~Pato, L.~Baudis, G.~Bertone, R.~Ruiz~de Austri, L.~E. Strigari, et~al.,
  {Complementarity of Dark Matter Direct Detection Targets}, Phys.Rev. D83
  (2011) 083505.
\newblock \href {http://arxiv.org/abs/1012.3458} {\path{arXiv:1012.3458}},
  \href {http://dx.doi.org/10.1103/PhysRevD.83.083505}
  {\path{doi:10.1103/PhysRevD.83.083505}}.

\bibitem{Newstead:2013pea}
J.~L. Newstead, T.~D. Jacques, L.~M. Krauss, J.~B. Dent, F.~Ferrer, {The
  Scientific Reach of Multi-Ton Scale Dark Matter Direct Detection
  Experiments}, Phys.Rev. D88 (2013) 076011.
\newblock \href {http://arxiv.org/abs/1306.3244} {\path{arXiv:1306.3244}},
  \href {http://dx.doi.org/10.1103/PhysRevD.88.076011}
  {\path{doi:10.1103/PhysRevD.88.076011}}.

\bibitem{Baudis:2012ig}
L.~Baudis, {Direct dark matter detection: the next decade}, Phys.Dark Univ. 1
  (2012) 94--108.
\newblock \href {http://arxiv.org/abs/1211.7222} {\path{arXiv:1211.7222}},
  \href {http://dx.doi.org/10.1016/j.dark.2012.10.006}
  {\path{doi:10.1016/j.dark.2012.10.006}}.

\end{thebibliography}

\end{document}